\documentclass[aps, prb, twocolumn, superscriptaddress]{revtex4-2}
\usepackage[colorlinks=true,citecolor=blue,linkcolor=blue,urlcolor=blue]{hyperref}
\usepackage{graphicx, nicefrac, textcomp}
\usepackage[normalem]{ulem}
\usepackage{changes}
\usepackage{chngcntr}
\usepackage{amssymb}
\usepackage{amsmath}

\begin{document}

\title{Evolution of plasmon excitations across the phase diagram of the cuprate superconductor La$_{2-x}$Sr$_{x}$CuO$_4$}

\author{M.~Hepting}
\email[]{hepting@fkf.mpg.de}
\affiliation{Max-Planck-Institute for Solid State Research, Heisenbergstra{\ss}e 1, 70569 Stuttgart, Germany}
\author{T.~D.~Boyko}
\affiliation{Canadian Light Source, Saskatoon, Saskatchewan S7N 2V3, Canada}
\author{V.~Zimmermann}
\affiliation{Max-Planck-Institute for Solid State Research, Heisenbergstra{\ss}e 1, 70569 Stuttgart, Germany}
\author{M.~Bejas}
\affiliation{Facultad de Ciencias Exactas, Ingenier\'{\i}a y Agrimensura and Instituto de F\'{\i}sica de Rosario (UNR-CONICET), Avenida Pellegrini 250, 2000 Rosario, Argentina}
\author{Y.~E.~Suyolcu}
\affiliation{Max-Planck-Institute for Solid State Research, Heisenbergstra{\ss}e 1, 70569 Stuttgart, Germany}
\author{P.~Puphal}
\affiliation{Max-Planck-Institute for Solid State Research, Heisenbergstra{\ss}e 1, 70569 Stuttgart, Germany}
\author{R.~J.~Green}
\affiliation{Department of Physics \& Engineering Physics, University of Saskatchewan, Saskatoon, Saskatchewan, Canada}
\affiliation{Stewart Blusson Quantum Matter Institute, University of British Columbia, Vancouver, British Columbia V6T 1Z1, Canada}
\author{L.~Zinni}
\affiliation{Facultad de Ciencias Exactas, Ingenier\'{\i}a y Agrimensura (UNR), Avenida Pellegrini 250, 2000 Rosario, Argentina}
\author{J.~Kim}
\affiliation{Advanced Photon Source, Argonne National Laboratory, Argonne, Illinois 60439, USA}
\author{D.~Casa}
\affiliation{Advanced Photon Source, Argonne National Laboratory, Argonne, Illinois 60439, USA}
\author{M.~H.~Upton}
\affiliation{Advanced Photon Source, Argonne National Laboratory, Argonne, Illinois 60439, USA}
\author{D.~Wong}
\affiliation{Helmholtz-Zentrum Berlin f{\"u}r Materialien und Energie, Hahn-Meitner-Platz 1, D-14109 Berlin, Germany}
\author{C.~Schulz}
\affiliation{Helmholtz-Zentrum Berlin f{\"u}r Materialien und Energie, Hahn-Meitner-Platz 1, D-14109 Berlin, Germany}
\author{M.~Bartkowiak}
\affiliation{Helmholtz-Zentrum Berlin f{\"u}r Materialien und Energie, Hahn-Meitner-Platz 1, D-14109 Berlin, Germany}
\author{K.~Habicht}
\affiliation{Helmholtz-Zentrum Berlin f{\"u}r Materialien und Energie, Hahn-Meitner-Platz 1, D-14109 Berlin, Germany}
\affiliation{Institute of Physics and Astronomy, University of Potsdam, Karl-Liebknecht-Stra{\ss}e 24/25, D-14476 Potsdam, Germany}
\author{E.~Pomjakushina}
\affiliation{Laboratory for Multiscale Materials Experiments (LMX), Paul Scherrer Institute (PSI), CH-5232 Villigen, Switzerland}
\author{G.~Cristiani}
\affiliation{Max-Planck-Institute for Solid State Research, Heisenbergstra{\ss}e 1, 70569 Stuttgart, Germany}
\author{G.~Logvenov}
\affiliation{Max-Planck-Institute for Solid State Research, Heisenbergstra{\ss}e 1, 70569 Stuttgart, Germany}
\author{M.~Minola}
\affiliation{Max-Planck-Institute for Solid State Research, Heisenbergstra{\ss}e 1, 70569 Stuttgart, Germany}
\author{H.~Yamase}
\affiliation{International Center of Materials Nanoarchitectonics, National Institute for Materials Science, Tsukuba 305-0047, Japan}
\author{A.~Greco}
\email[]{agreco@fceia.unr.edu.ar}
\affiliation{Facultad de Ciencias Exactas, Ingenier\'{\i}a y Agrimensura and Instituto de F\'{\i}sica de Rosario (UNR-CONICET), Avenida Pellegrini 250, 2000 Rosario, Argentina}
\author{B.~Keimer}
\affiliation{Max-Planck-Institute for Solid State Research, Heisenbergstra{\ss}e 1, 70569 Stuttgart, Germany}

\begin{abstract}
We use resonant inelastic x-ray scattering (RIXS) at the O $K$- and Cu $K$-edges to investigate the doping- and temperature dependence of low-energy plasmon excitations in La$_{2-x}$Sr$_{x}$CuO$_4$. We observe a monotonic increase of the energy scale of the plasmons with increasing doping $x$ in the underdoped regime, whereas a saturation occurs above optimal doping $x \gtrsim 0.16$ and persists at least up to $x = 0.4$. Furthermore, we find that the plasmon excitations show only a marginal temperature dependence, and possible effects due to the superconducting transition and the onset of strange metal behavior are either absent or below the detection limit of our experiment. Taking into account the strongly correlated character of the cuprates, we show that layered $t$-$J$-$V$ model calculations accurately capture the increase of the plasmon energy in the underdoped regime. However, the computed plasmon energy continues to increase even for doping levels above $x \gtrsim 0.16$, which is distinct from the experimentally observed saturation, and reaches a broad maximum around $x = 0.55$. We discuss whether possible lattice disorder in overdoped samples, a renormalization of the electronic correlation strength at high dopings, or an increasing relevance of non-planar Cu and O orbitals could be responsible for the discrepancy between experiment and theory for doping levels above $x = 0.16$.

\end{abstract}

\date{\today}

\maketitle


\section{Introduction}
\label{sec:introduction}

In spite of intense experimental and theoretical research efforts, superconducting cuprates have retained an enigmatic character for more than three decades. On the one hand, it is well known that their high-temperature superconductivity emerges when charge carriers are doped into the CuO$_2$ planes, suppressing the long-range antiferromagnetic order that is prevalent in the Mott insulating parent compounds  \cite{Keimer2015,Scalapino2012,Armitage2010,Lee2006}. For instance, in the prototypical cuprate La$_{2-x}$Sr$_{x}$CuO$_4$, the doping of $p \approx 0.07$ holes per CuO$_2$ unit is sufficient to stabilize superconductivity \cite{Takagi1992}, and the highest transition temperature $T_c$ is realized for $p \approx 0.16$.
On the other hand, a consensus on the microscopic mechanism mediating the superconductivity in cuprates is still lacking. This elusive nature of superconductivity in cuprates is mostly rooted in the fact that---in contrast to conventional superconductors---cuprates are strongly correlated systems and their Cooper pairs do not condense from a uniform and well-understood metallic state. Instead, the superconducting dome as a function of charge carrier doping is embedded in a variety of other enigmatic phases that also exhibit a strongly correlated character, and either promote or compete with superconductivity in a nontrivial fashion. Most prominently, these phases comprise the pseudogap, spin and charge order, as well as the strange metal regime \cite{Keimer2015}.   

Insights into the electronic structure and the dynamics of doped charge carriers in these diverse phases can be gained by various spectroscopic techniques. For instance, x-ray absorption spectroscopy (XAS) and electron-energy loss spectroscopy (EELS) at the O $K$-edge or Cu $L$-edge have revealed how doped holes distribute within the CuO$_2$ planes, that is, they reside within the hybridized Cu $3d_{x^2-y^2}$ and O 2$p_{x,y}$ orbitals, with a dominant influence of the latter orbitals \cite{Nucker1988,Chen1991,Pellegrin1993,Haruta2018}. This behavior of the doped holes is commonly translated into a low-energy effective model where the quasiparticles are Zhang-Rice singlets (ZRS) \cite{Tjeng1997,Brookes2015}, which correspond to plaquettes of holes and oxygen ions around a Cu ion, taking on the same role as fully occupied or empty sites in an effective single-band Hubbard model \cite{Zhang1988}. The single-band approach is also employed in the $t-J$ model \cite{Dagotto1994,Ogata2008,Hybertsen1990,Aligia1994}, which additionally discards all doubly occupied states. 
Nevertheless, an accurate description of specific properties of cuprates, such as $pd$ charge-transfer related effects, requires the explicit consideration of all three planar orbitals (Cu $3d_{x^2-y^2}$ and O 2$p_{x,y}$) \cite{Emery1987,Varma1997,Weber2008,Weber2012,Chen2013,Chen2013a,Ebrahimnejad2014,Ebrahimnejad2016,Adolphs2016,Kung2016}, and in some cases also non-planar orbitals \cite{Bianconi1988,Grilli1990,Romberg1990,Feiner1992,Chen1992,Pavarini2001,Wang2010,Weber2010,Sakakibara2010,Hozoi2011,Sakurai2011,Sakakibara2012,Matt2018,Kramer2019,Jiang2020}.  

A different technique for investigating the charge dynamics of cuprates, both in the normal and superconducting state, is infrared/optical spectroscopy \cite{Basov2005,Suzuki1989,Uchida1991}. Yet, in contrast to EELS, optical spectroscopy is a probe that is essentially limited to the center of the Brillouin zone (BZ), due to the small momentum of optical photons. 

Recently, resonant inelastic x-ray scattering (RIXS) \cite{Ament2011} has emerged as a versatile tool to study electronic and magnetic excitations in cuprates, providing both momentum resolution and a relatively high energy resolution. Along these lines, soft x-ray RIXS at the Cu $L_3$ and O $K$-edge was employed to investigate collective spin excitations \cite{Letacon2011,Dean2013,Bisogni2012a,Bisogni2012b}, and hard x-ray RIXS at the Cu $K$-edge was used to probe electronic inter- and intraband transitions \cite{Kim2002,Ishii2005a,Ishii2005b,Wakimoto2005,Ellis2011,Wakimoto2013}. Moreover, collective charge excitations were detected in Cu $L_3$ and O $K$-edge RIXS experiments, which were attributed to acoustic plasmons \cite{Hepting2018,Lin2019,Nag2020,Singh2022}. More recently, a high-resolution Cu $L_3$-edge RIXS study revealed that these low-energy plasmon branches are not strictly acoustic, but they exhibit an energy gap at the two-dimensional BZ center \cite{Hepting2022}. 

Plasmons are a fundamental collective excitation of the charge carrier density in metallic systems, mediated by long-range Coulomb interactions. An isotropic three-dimensional metal usually exhibits only one optical plasmon branch, while systems with stacked conducting planes that interact via poorly-screened interlayer Coulomb interactions exhibit a set of acoustic plasmon branches that disperse almost linearly towards zero energy at the BZ center \cite{Grecu1973,fetter1974,Grecu1975,Gabriele2022,Boyd2022}. However, in the presence of single-electron tunneling between the planes, the latter branches are not strictly acoustic, but exhibit a gap at the BZ center, which is proportional to the interlayer hopping integral $t_z$ \cite{Hepting2022,Greco2016}. In doped cuprates, which are composed of stacked conducting CuO$_2$ planes separated by dielectric spacer layers, early transmission EELS and optical spectroscopy experiments already detected the optical plasmon, which manifests itself as a plasma edge in reflectance spectra and a peak in the loss function around 1 eV \cite{Bozovic1990,Fink2001}. Yet, direct evidence of the acoustic-like plasmon branches was only provided later by high-resolution RIXS with polarization analysis \cite{Hepting2018} and the capacity to independently vary the in- 
and out-of-plane 
momentum transfer to probe the distinct plasmon dispersion along different directions in the BZ \cite{Hepting2018,Nag2020}. The plasmon dispersion is accurately captured by a large-$N$ theory of the layered $t$-$J$-$V$ model \cite{Greco2016,Greco2019,Greco2020,Nag2020,Hepting2022}, which intrinsically accounts for the strongly correlated character of cuprates. Other methods to describe the plasmon dispersion in cuprates include the random phase approximation (RPA) \cite{Markiewicz2007a,Kresin1988}, a combination of determinant quantum Monte Carlo (DQMC) and RPA in a layered Hubbard model \cite{Hepting2018}, and an extended variational wave function approach  \cite{Fidrysiak2021}.

In a broader context, low-energy plasmons in cuprates attracted considerable attention already before their detection with RIXS \cite{Bozovic1990,Fink2001}, as they were suggested to play a role in the superconducting pairing \cite{Kresin1988}, and their evolution across the superconducting transition could reflect possible savings of kinetic energy of the charge carriers \cite{Levallois2016}. Moreover, low-energy plasmon-phonon modes \cite{Falter1994,Bauer2009} were proposed to mediate superconductivity, or contribute constructively to the high $T_c$ of cuprates \cite{Bill2003}. In addition, the evolution of plasmon excitations across the phase diagram of cuprates might encode critical information about the charge carrier dynamics in the normal state. Specifically, a debate is ongoing about the nature of the charge carriers in the strange metal state \cite{Varma1989,Phillips2022}, which is characterized by a linear-in-temperature resistivity \cite{Varma2020}. Several theories suggest that the standard Landau quasiparticle description for the charge carriers breaks down in this regime \cite{Zaanen2019,Hartnoll2021,Chowdhury2022}, inhibiting the emergence of collective charge excitations. As a consequence, strange metals could hamper the propagation of plasmons, while they rapidly decay into a quantum critical continuum \cite{Romero2019}. Indeed, the lack of well-defined plasmon peaks in EELS data acquired in a reflection geometry was interpreted as a sign of such an anomalous plasmon damping in overdoped cuprates, \cite{Mitrano2019,Husain2019}, which is still discussed controversially \cite{Fink2021,Husain2021}.

In this work, we use soft x-ray RIXS at the O $K$-edge and hard x-ray RIXS at the Cu $K$-edge to comprehensively map the evolution of low-energy plasmon excitations across the phase diagram of La$_{2-x}$Sr$_{x}$CuO$_4$. Whereas the doping dependence of plasmon excitations in electron-doped cuprates was addressed in earlier studies \cite{Hepting2018,Lin2019}, the evolution of the plasmon as a function of hole-doping has not been systematically investigated by RIXS, to the best of our knowledge. We observe plasmon excitations for several Sr substitution levels from $x = 0.05$ to 0.4 and at various temperatures, including the strange metal regime around $x = 0.2$, where previous works proposed the decay of the plasmon into an energy- and momentum-independent continuum \cite{Mitrano2019,Husain2019,Romero2019}. We model the doping dependence of the plasmon energy in the framework of layered $t$-$J$-$V$ model calculations, which capture the experimentally observed trends but show a deviation for dopings above $x \gtrsim 0.16$ (optimal doping). We discuss the presence of disorder, a renormalization of the electronic correlation strength in the overdoped regime, and an increasing relevance of non-planar Cu and O orbitals as possible origins for this discrepancy.

\section{Results}
\label{sec:results}

\subsection{Electronic transport}

\begin{figure}[tb]
\includegraphics[width=.99\columnwidth]{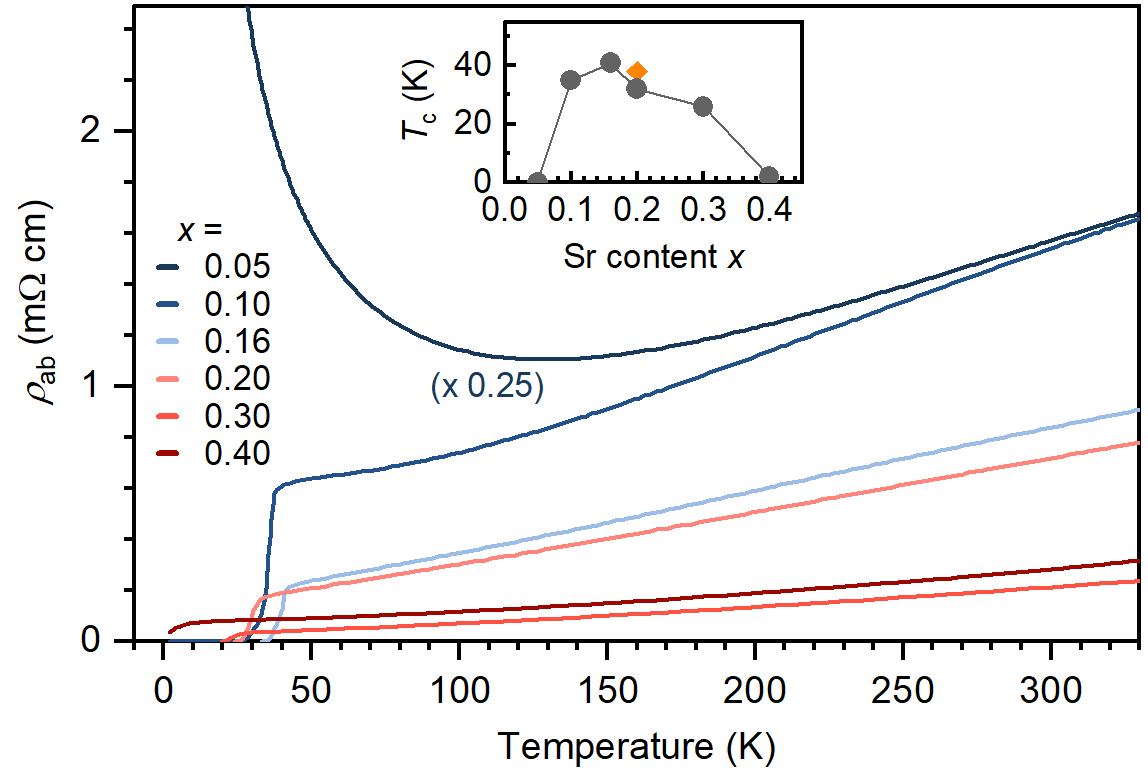}
\caption{In-plane resistivity $\rho_{ab}$ of La$_{2-x}$Sr$_{x}$CuO$_4$ films with various hole-doping levels $x$. The $x$ = 0.05 curve is scaled by a factor 0.25. The gray symbols in the inset correspond to the superconducting transition temperatures $T_c$ of the films. The orange symbol indicates the $T_c$ of the La$_{1.8}$Sr$_{0.2}$CuO$_4$ single-crystal.  
}
\label{transport}
\end{figure}

Thin films of La$_{2-x}$Sr$_{x}$CuO$_4$ were synthesized by ozone-assisted molecular beam epitaxy (MBE) (see Appendix~\ref{sec:experiment} for details). Our film with the lowest doping ($x$ = 0.05) is non-superconducting and becomes insulating at low temperatures [Fig.~\ref{transport}], whereas the subsequent doping level $x$ = 0.1 shows a clear superconducting transition. This is consistent with previous La$_{2-x}$Sr$_{x}$CuO$_4$ thin film studies \cite{Sato2000,Ando2004a,Bozovic2016} and bulk measurements, where the onset of the superconducting dome is situated around $x \sim 0.07$ \cite{Takagi1992}. Note that the long-range antiferromagnetic (AFM) order, which is prevalent in the parent compound La$_{2}$CuO$_4$, vanishes already around $x \sim 0.016$ \cite{Takagi1989}. For higher dopings, spin and charge stripe ordered states emerge and are most pronounced around $x \sim 0.125$. In the bulk, a low-temperature-tetragonal (LTT) and low-temperature less orthorhombic (LTLO) structural phase transition also occur in this doping regime \cite{Kastner1998}. In thin epitaxial La$_{2-x}$Sr$_{x}$CuO$_4$ thin films, however, these structural transitions are suppressed \cite{Ando2004a}.

As expected, we detect the highest superconducting transition of our film series for the optimal doping $x$ = 0.16, that is, $T_c$ = 41~K. The $T_c$ decreases slightly in the subsequent film with $x$ = 0.2, where the electrical resistivity varies almost perfectly linear as a function of temperature in the normal state at least up to 300 K [Fig.~\ref{transport}]. Such $T$-linear behavior is commonly regarded as a key manifestation of the strange metal phenomenology \cite{Zaanen2019,Phillips2022,Hartnoll2021}. Previous transport studies on La$_{2-x}$Sr$_{x}$CuO$_4$ found that the $T$-linear behavior is restored at low temperatures when the superconducting transition is suppressed in strong magnetic fields, and that the strange metal phase emanates in a fan-like shape from a putative quantum critical point at $p^* \sim 0.19$ \cite{Cooper2009}.
In addition to a series of thin films, we have synthesized a single-crystal with a doping concentration close to $x$ = 0.2. The temperature dependence of the transport and the $T_c$ of the single-crystal are closely similar to the corresponding film, and will be discussed in more detail below in the context of the Cu $K$-edge RIXS measurements. 

For the overdoped $x$ = 0.3 film, we observe a decrease of $T_c$ to 26 K. The highly overdoped $x$ = 0.4 film exhibits an onset of a superconducting transition around 5 K. This behavior in the latter two samples differs from reports on bulk La$_{2-x}$Sr$_{x}$CuO$_4$, where the superconducting dome typically terminates around $x \sim 0.25$ \cite{Takagi1989}, and from prior thin film studies that reported only a slightly extended dome \cite{Sato2000}. In the case of our films, we attribute the substantially extended superconducting dome to the highly oxidizing growth atmosphere \cite{Sederholm2021} and the stabilizing effect of the epitaxial strain of the substrate, which can possibly enhance the solubility limit of Sr \cite{Tanaka1994} and alleviate the formation of oxygen vacancies \cite{Kim2017,Wang2022}. A similar trend was recently observed in ozone-assisted MBE grown La$_{2-x}$Ca$_{x}$CuO$_4$ films, where the superconducting transition persisted for Ca-substitutions as high as $x$ = 0.5 \cite{Kim2021}.

\begin{figure*}[tb]
\includegraphics[width=2.0\columnwidth]{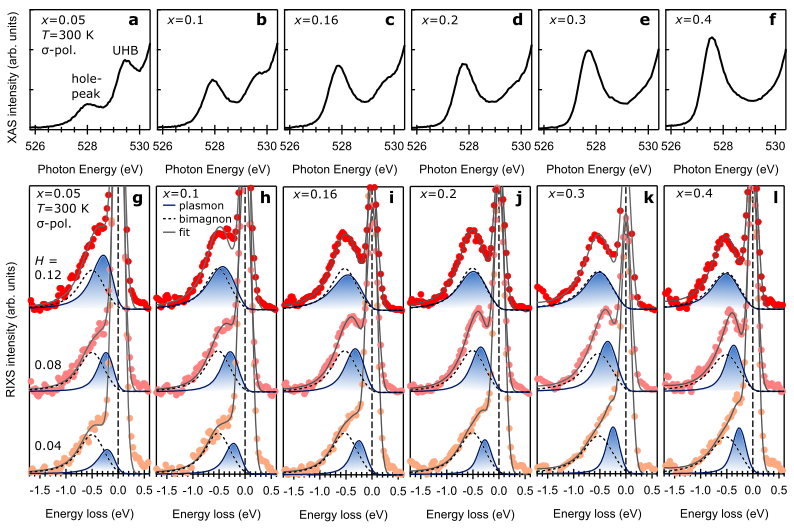}
\caption{(a)-(f) O $K$-edge XAS of La$_{2-x}$Sr$_{x}$CuO$_4$ films with various hole-doping levels $x$. The spectra were collected at $T$ =300 K with $\sigma$-polarized photons at an incident angle $\theta = 35^\circ$. The peak associated with the doped holes and the upper Hubbard band (UHB) peak are indicated. (g)-(l) O $K$-edge RIXS of the La$_{2-x}$Sr$_{x}$CuO$_4$ films, for various momenta along the $H$ direction, while $K = 0$ and $L^* \approx 0.37$. Orange symbols correspond to $H$ = 0.04, light red symbols to $H$ = 0.08, and red symbols to $H$ = 0.12. The spectra were taken with $\sigma$-polarized photons at energies tuned to the hole-peak in the corresponding XAS. A high-energy background was subtracted from the spectra (see Appendix~\ref{app:fit}). The solid gray lines are fits, which include the dispersive plasmon (blue shaded peak) and the non-dispersive bimagnon (dashed black peak). The other contributions to the fit are omitted for clarity (for details of the fitting procedure see Appendix~\ref{app:fit}).
Curves for different momenta are offset in the vertical direction for clarity.  
}
\label{LSCO_OK}
\end{figure*}

\subsection{X-ray absorption at the O $K$-edge}

The XAS signal across the near-edge fine structure of the O $K$-edge contains hallmark fingerprints of the electronic structure of the cuprates \cite{Chen1991,Chen1992,Pellegrin1993,Chen2013,Brookes2015,Wang2010}. Figures~\ref{LSCO_OK}a-f display the O $K$-edge XAS of the La$_{2-x}$Sr$_{x}$CuO$_4$ films with dopings between $x$ = 0.05 and 0.4. The peak near 529.5~eV is particularly pronounced for low doping levels and corresponds to the upper Hubbard band (UHB), which arises as the on-site part of the Coulomb repulsion $U$ splits the Cu 3$d$ band into a fully occupied lower Hubbard band (LHB) and the empty UHB. The Cu 3$d$ band-derived features are visible in the O $K$-edge XAS due to the strong 2$p$-3$d$ hybridization between O and Cu.
The peak in the XAS at lower energies ($\sim$ 528~eV) typically emerges in hole-doped cuprates \cite{Abbamonte2005}, and is associated with the ZRS states \cite{Brookes2015}. 
With increasing hole density, spectral weight is transferred from the UHB peak to the hole-peak. 
Notably, we observe that with increasing Sr-substitution the intensity of the hole-peak increases continuously and the peak energy shifts to lower energies, even up to $x$ = 0.4. This trend is consistent with the notion of a continuous creation of new doped hole states and that the ZRS stays intact up to highest doping levels \cite{Chen2013,Brookes2015}.

\subsection{RIXS at the O $K$-edge}

\begin{figure}[tb]
\includegraphics[width=1.0\columnwidth]{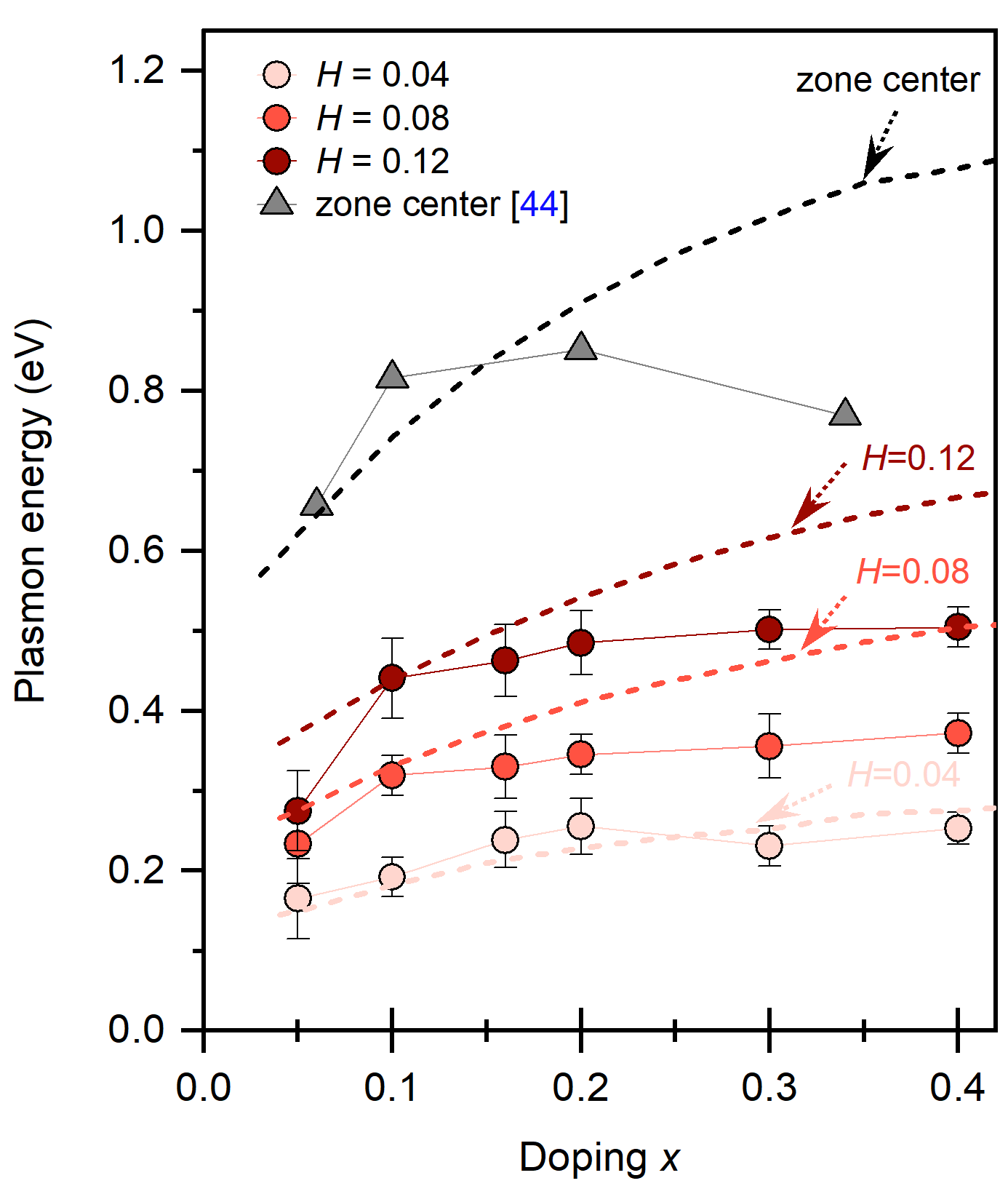}
\caption{Doping-dependence of the plasmon energy. The circles are the plasmon energies extracted from fits of the O $K$-edge RIXS spectra of the La$_{2-x}$Sr$_{x}$CuO$_4$ films in Figs.~\ref{LSCO_OK}g-l. The energies are grouped according to the different in-plane momentum transfer $H$ (light to dark red). The gray triangles correspond the plasmon energy at the three-dimensional Brillouin zone center $H=K=L=0$ (optical plasmon), reported for La$_{2-x}$Sr$_{x}$CuO$_4$ single-crystals in Ref.~\onlinecite{Uchida1991}. The dashed lines are computed plasmon energies obtained from layered $t$-$J$-$V$ model calculations. The labels next to the dashed arrows indicate the momenta for which the calculations were performed.  
}
\label{dopingtheo}
\end{figure}

In order to gain deeper insights into the dynamics of the doped holes and the ZRS state for each doping level, we measured RIXS with the incident photon energy tuned to the maximum of the hole-peak in the corresponding XAS from Figs.~\ref{LSCO_OK}a-f. In the following, we denote the momentum transfer by $(H,K,L)$ in reciprocal lattice units $(2\pi/a, 2\pi/b, 2\pi/c)$, where $a$, $b$, and $c$ are the lattice constants of La$_{2-x}$Sr$_{x}$CuO$_4$. However, the crystallographic unit cell of La$_{2-x}$Sr$_{x}$CuO$_4$ contains two CuO$_2$ planes, whereas only the distance $d = c/2$ between adjacent CuO$_2$ planes is relevant for the periodicity of the plasmon dispersion \cite{Hepting2018}. Hence, we utilize the index $L^* = L/2$ instead.

The obtained RIXS spectra for in-plane momenta $H$ = 0.04, 0.08, and 0.12 are displayed in Figs.~\ref{LSCO_OK}g-l. 
The out-of-plane momentum transfer of the spectra is on average $L^* \sim$ 0.37 (see Appendix~\ref{sec:experiment} for details). A broad background contribution in the RIXS signal from fluorescence and $dd$-excitations \cite{Bisogni2012a} is already subtracted from the spectra in Figs.~\ref{LSCO_OK}g-l. The raw spectra without background subtraction and details about the fitting procedure can be found in Appendix~\ref{app:fit}.

In Figs.~\ref{LSCO_OK}g-l, the spectra are decomposed into several components, including the elastic line centered at zero-energy loss, a non-dispersive feature around 510 meV, and a dispersive peak that evolves between 200 and 500 meV.  Along the lines of previous O $K$-edge RIXS studies, we assign the $\sim$510 meV peak (dashed black line in Figs.~\ref{LSCO_OK}g-l) to a non-dispersive bimagnon excitation \cite{Bisogni2012a,Bisogni2012b,Nag2020,Singh2022}. Note that long-range AFM order is absent for the investigated doping levels, hence, this excitation corresponds strictly speaking to a bi-paramagnon \cite{Letacon2011,Dean2013} and not a bimagnon.  The essentially unchanged energy scale of our observed non-dispersive feature is consistent with the evolution of paramagnon excitations in Cu $L$-edge RIXS experiments on La$_{2-x}$Sr$_{x}$CuO$_4$ films with doping levels between $x$=0.1 and 0.4 \cite{Dean2013}. In the following, we will focus on the dispersive peak (solid blue line in Figs.~\ref{LSCO_OK}g-l), which we attribute to a low-energy plasmon excitation. Notably, the bimagnon and the plasmon peak overlap strongly for momenta between $H$ = 0.04 and 0.12 [Figs.~\ref{LSCO_OK}g-l], which precluded the identification of two distinct peaks in early RIXS experiments on La$_{2-x}$Sr$_{x}$CuO$_4$ \cite{Bisogni2012b,Ishii2017}. Nevertheless, recent O $K$-edge RIXS studies with improved energy resolution \cite{Nag2020,Singh2022} clearly discerned the two excitations and determined the plasmon energy as a function of momentum transfer $H$. Similarly, we extract the energy of the dispersive plasmon peak from our fits, while keeping the energy of the bimagnon peak fixed. For La$_{1.84}$Sr$_{0.16}$CuO$_4$, we obtain closely similar plasmon energies [Fig.~\ref{LSCO_OK}i and Fig.~\ref{dopingtheo}] as reported in Ref.~\onlinecite{Nag2020} for the same doping (see Appendix~\ref{app:literature}). Furthermore, our extracted plasmon energies for La$_{1.9}$Sr$_{0.1}$CuO$_4$ [Fig.~\ref{LSCO_OK}h and Fig.~\ref{dopingtheo}] are not far from those reported for La$_{1.88}$Sr$_{0.12}$CuO$_4$ in Ref.~\onlinecite{Singh2022} (see also Appendix~\ref{app:literature}).  
Yet, to the best of our knowledge, the plasmon energy for other doping levels of La$_{2-x}$Sr$_{x}$CuO$_4$ has not yet been reported. Thus, we closely inspect Figs.~\ref{LSCO_OK}g-l in the following to extract a possible systematic trend in the doping-dependence of the plasmon excitation.

Figure~\ref{LSCO_OK}g shows the RIXS spectra for the lightly doped case $x = 0.05$, where our fits reveal that the broad shoulder next to the elastic line contains the bimagnon and the plasmon excitation, similar to the $x = 0.1$ and 0.16 cases [Figs.~\ref{LSCO_OK}h,i]. However, the energy scale and the dispersion of the plasmon for $x = 0.05$ are substantially smaller than for $x = 0.1$ and 0.16. This difference is also visible in Fig.~\ref{dopingtheo}, which plots the doping dependence of the plasmon energy for each momentum $H$, suggesting a monotonic increase of the energy as a function of $x$.
An increase of the plasmon energy with increasing doping concentration was also observed in RIXS experiments on the electron-doped cuprate La$_{2-x}$Ce$_{x}$CuO$_4$ \cite{Hepting2018,Lin2019}. Note that this trend does not depend on the details of our fitting procedure, such as the exact treatment of the bimagnon peak. For instance, a fit that captures both the plasmon and the bimagnon feature by a single broad peak function still yields an increase of the peak energy with increasing doping concentration (see Appendix~\ref{app:fit}). 

For dopings beyond $x = 0.16$, we find that the plasmon energy does not continue to grow, but instead tends to saturate for $x = 0.2$, 0.3, and 0.4 [Figs.~\ref{LSCO_OK}j-l and Fig.~\ref{dopingtheo}]. The onset of a flattening of the plasmon energy increase in the overdoped regime was already noticed for electron-doped La$_{2-x}$Ce$_{x}$CuO$_4$ and attributed to a variation in the band dispersion and Fermi surface reconstruction \cite{Hepting2018}, or strong correlations \cite{Lin2019}. However, the investigated range for La$_{2-x}$Ce$_{x}$CuO$_4$ was between $x = 0.11$ and 0.18, whereas our present study on La$_{2-x}$Sr$_{x}$CuO$_4$ extends over a much wider range from $x = 0.05$ to 0.4, which can allow for more robust conclusions. 

Besides the systematic evolution of the plasmon energy [Fig.~\ref{dopingtheo}], we find that the integrated intensity of the plasmon peak changes only marginally, especially in the overdoped regime [Figs.~\ref{LSCO_OK}j-l]. On the other hand, a small decrease in the linewidth of the plasmon for high dopings might be discernible in Figs.~\ref{LSCO_OK}j-l. Yet, because of the strong overlap with the bimagnon, the statistical error of the integrated intensities and linewidths extracted from the fits are large and we refrain from further discussions of these two quantities.

Nonetheless, the RIXS data in Figs.~\ref{LSCO_OK}g-l clearly demonstrate that plasmon excitations pervade the cuprate phase diagram at least from $x = 0.05$ to 0.4 and for momenta as large as $H$ = 0.12. In particular, the unaffected emergence of plasmon quasiparticles in the overdoped regime raises the question, whether such an observation is compatible with previous proposals based on reflection EELS experiments \cite{Mitrano2019,Husain2019,Husain2021} that suggested the vanishing of coherent plasmon excitations specifically in the strange metal regime of cuprates. Hence, in the following we focus on the doping level $x = 0.2$, which is situated within the strange metal region in the La$_{2-x}$Sr$_{x}$CuO$_4$ phase diagram  \cite{Cooper2009,Ando2004a}.

\subsection{RIXS at the Cu $K$-edge}

\begin{figure}[tb]
\includegraphics[width=.95\columnwidth]{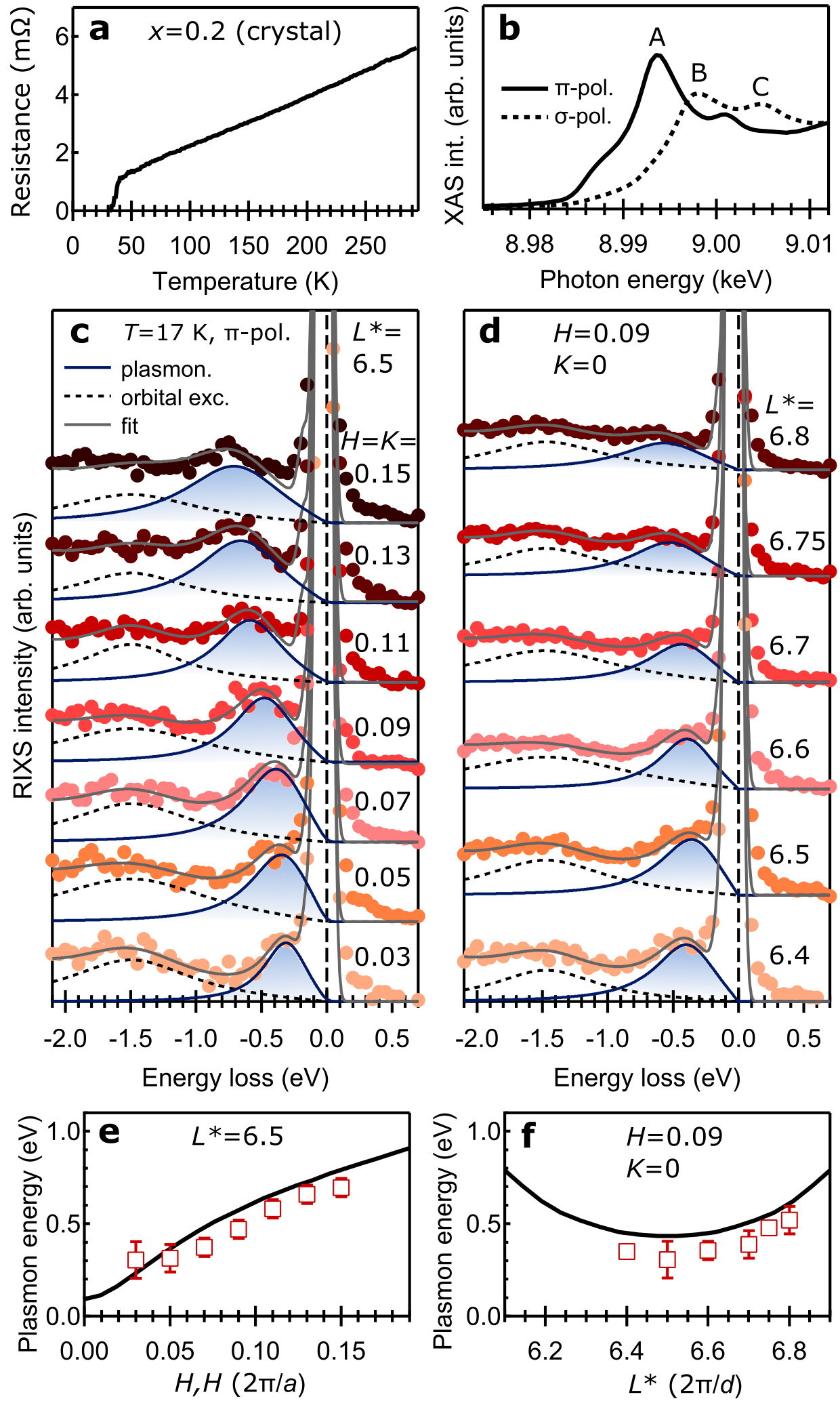}
\caption{Cu $K$-edge RIXS of the La$_{1.8}$Sr$_{0.2}$CuO$_4$ single-crystal. (a) Electrical transport of the crystal. (b) XAS across the Cu $K$-edge measured with $\pi$- (solid line) and $\sigma$-polarized (dashed line) incident photons. (c) Cu $K$-edge RIXS spectra for various in-plane momenta along the diagonal direction ($H = K$), while the out-of-plane momentum is fixed to $L^* = 6.5$. The fit (gray line) to the experimental data (filled symbols) includes the plasmon peak (blue shaded peak profile), an orbital excitation (dashed black line), and other contributions (not shown here) that are described in Appendix~\ref{app:fit}. Curves for different momenta are offset in the vertical direction for clarity. (d) RIXS spectra for momenta along the $L^*$ direction, while $H=0.09$ and $K=0$ are fixed. (e,f) Plasmon energy (red squares) extracted from the fits (e) along the diagonal direction and (f) along the $L^*$ direction. The solid black lines are the corresponding plasmon dispersions computed in the $t$-$J$-$V$ model. 
}
\label{CuK}
\end{figure}

To corroborate the existence of plasmon excitations in the strange metal regime and to obtain complementary information, we use RIXS at a different absorption edge, that is, the Cu $K$-edge. An advantage of the high energy of the hard x-ray photons tuned to the Cu $K$-edge is the wide range of reciprocal space that becomes accessible for RIXS. However, due to the increased penetration depth of the photons, we carry out the RIXS measurements on a bulk single-crystal instead of the La$_{1.8}$Sr$_{0.2}$CuO$_4$ thin film that was used for the O $K$-edge experiment above. Our single-crystal exhibits a $T_c$ of approximately 38\,K and a similar $T$-linear transport behavior [Fig.~\ref{CuK}a] as the La$_{1.8}$Sr$_{0.2}$CuO$_4$ thin film [Fig.~\ref{transport}], indicating that the sample is excellently suitable for complementary RIXS measurements.

Figure~\ref{CuK}b shows the XAS spectra of the La$_{1.8}$Sr$_{0.2}$CuO$_4$ single-crystal across the Cu $K$-edge measured with $\pi$- and $\sigma$-polarized incident photons, respectively. The main absorption line for the former polarization is centered at 8993.7\,eV (peak A), while the latter polarization yields pronounced peaks around 8998.1\,eV (peak B) and 9004.9\,eV (peak C). The shapes of our XAS spectra are in close agreement with those of overdoped La$_{1.7}$Sr$_{0.3}$CuO$_4$ in Ref.~\onlinecite{Wakimoto2013}. A relative shift of the absolute energy scale of approximately 1\,eV in Ref.~\onlinecite{Wakimoto2013} is likely due to a different energy calibration.

In order to identify the resonance energy where the emergence of the plasmon excitation is expected in RIXS, we briefly recall previous assignments of the Cu $K$-edge XAS features of cuprates. Note that low-energy plasmons were not observed in previous Cu $K$-edge RIXS experiments on hole-doped cuprates, whereas recent Cu $K$-edge RIXS studies on electron-doped cuprates revealed a low-energy excitation that is compatible with the typical plasmon dispersion \cite{Ishii2019}. Yet, the XAS of electron-doped cuprates includes several distinctions from the hole-doped counterparts \cite{Ishii2005a}. In hole-doped La$_{2-x}$Sr$_{x}$CuO$_4$, the XAS peak A was attributed to well-screened intermediate states, and previous RIXS studies detected charge-transfer (CT) excitations around 4\,eV energy loss in resonance to this energy \cite{Wakimoto2005,Wakimoto2013,Kim2002}. For resonant energies in the region of peak B and C, which are associated with poorly screened states, molecular-orbital excitations around 7.5\,eV energy loss were observed with RIXS \cite{Wakimoto2005,Wakimoto2013,Kim2002}. However, both types of high-energy excitations are clearly not associated with low-energy plasmon modes. Instead, the plasmon excitation must be located in the CT gap, where indeed spectral weight below 3\,eV energy loss was reported, which showed a dependence on the in-plane momentum transfer \cite{Wakimoto2005,Ellis2011,Wakimoto2013}. This broad spectral weight, whose intensity resonates at the energy of peak A, was initially attributed to an interband excitation from a low-lying band below the Fermi energy to the ZRS band \cite{Ellis2011}. In contrast, an analysis of the spectral intensity specifically around 1\,eV in Ref.~\onlinecite{Wakimoto2013} suggested an intraband excitation as the origin, although this assignment is incompatible with theoretical calculations \cite{Jia2012}. Nonetheless, Ref.~\onlinecite{Wakimoto2013} pointed out that for incident energies around peak A, the RIXS intensity inside the CT gap contains an appreciable contribution from another type of excitation, possibly unresolved due to an insufficient energy resolution of $\Delta E \sim$ 350\,meV in the RIXS experiment \cite{Wakimoto2013}. 

Hence, we focus on incident energies in the region around peak A in our RIXS experiment, and use a much improved energy resolution of $\Delta E \sim$ 80\,meV. To reduce the strong elastic scattering at the Cu $K$-edge, we employ $\pi$-polarized photons and keep the scattering angle close to $90^{\circ}$ for all measurements. This condition is realized for out-of-plane momenta revolving around $L^* = 6.5$ and small in-plane momenta. Further details about the scattering geometry are given in Appendix~\ref{sec:experiment}. 

Figure~\ref{CuK}c shows our Cu $K$-edge RIXS spectra for different in-plane momenta along the diagonal direction with $H = K$, while the out-of-plane momentum is fixed to $L^* = 6.5$. The RIXS spectra were taken with an incident photon energy of 8993\,eV, which is slightly below the maximum of the XAS peak A [Fig.~\ref{CuK}b]. Below 2\,eV energy loss in the spectra, we resolve two prominent inelastic features (for details about the fitting procedure, see Appendix~\ref{app:fit}). The feature centered around 1.5\,eV energy loss (dashed black line in Fig.~\ref{CuK}c) shows only little or no dispersion as a function of momentum transfer, whereas the feature at lower energies (solid blue line in Fig.~\ref{CuK}c) disperses strongly for varied momentum transfer. Both features resonate within an incident energy interval of more than 1\,eV above and below the maximum of XAS peak A. Upon increasing the incident energy, we find that the inelastic feature that appears at 1.5\,eV for an incident energy of 8993\,eV moves to higher energy losses. In contrast, we observed that the low-energy feature exhibits a Raman-like character and remains essentially at the same energy loss for different incident energies (not shown here). 

Furthermore, the latter feature shows a distinct out-of-plane dispersion [Fig.~\ref{CuK}d], while the 1.5\,eV feature remains at a constant energy loss within the experimental error. These behaviors indicate that the 1.5\,eV feature likely corresponds to an inter- or intraorbital excitation, as proposed for the features detected in Refs.~\onlinecite{Wakimoto2005,Ellis2011,Wakimoto2013}. In contrast, the behaviors of our dispersive feature at lower energies are compatible with a plasmon excitation, in analogy to the characteristic in- and out-of-plane dispersion of the plasmon mode recently identified in an O $K$-edge RIXS study on optimally doped La$_{1.84}$Sr$_{0.16}$CuO$_4$ \cite{Nag2020}, as well as that of electron-doped La$_{2-x}$Ce$_{x}$CuO$_4$ in an Cu $L$-edge RIXS study \cite{Hepting2018}. Moreover, the energy range of the dispersion of the low-energy feature in the Cu $K$-edge RIXS spectra is comparable to the range of the plasmon energy of the $x$ = 0.2 film in the O $K$-edge RIXS spectra in Fig.~\ref{LSCO_OK}j and Fig.~\ref{dopingtheo}. However, we note that the O $K$-edge RIXS spectra are acquired along the axial $H$ direction for $L^* = 0.37$, whereas the Cu $K$-edge spectra in Fig.~\ref{CuK}c are along the diagonal direction ($H = K$) for $L^* = 6.5$, which prevents a direct comparison. 
We rule out an assignment of the low-energy mode to a bimagnon excitation, as magnetic excitations in layered cuprates exhibit a two-dimensional character \cite{Hepting2018}, which would be at odds with the out-of-plane momentum dependence of our low-energy mode in Fig.~\ref{CuK}d. Note that previous Cu $K$-edge RIXS studies detected a bimagnon excitation in (underdoped) La$_{2-x}$Sr$_{x}$CuO$_4$, with a maximum in the intensity at the in-plane momentum ($\pi$, 0) and vanishing intensity at (0, 0) \cite{Ellis2010}. On the other hand, it was suggested that the bimagnon intensity in O $K$-edge RIXS is maximal around (0, 0), as the two types of RIXS measurements are sensitive to different parts of the bimagnon continuum \cite{Bisogni2012a}. The observation of a bi-(para)magnon in our O $K$-edge RIXS data [Figs.~\ref{LSCO_OK}g-l] and its absence (or very low intensity) in our Cu $K$-edge spectra [Figs.~\ref{CuK}c,d] is therefore consistent with this notion.

For a definitive assignment of the dispersive mode in Figs.~\ref{CuK}c,d, we turn to layered $t$-$J$-$V$ model calculations (for details see Appendix~\ref{sec:theory}). Specifically, we determine the plasmon energy as the maximum of the computed  imaginary part of the charge susceptibility $\chi''(\textbf{q},\omega)$  \cite{Greco2016,Greco2019,Greco2020,Nag2020,Hepting2022} to model the dispersion observed in Fig.~\ref{CuK}c along the ($H$,$H$) direction for fixed $L^* = 6.5$, as well as the dispersion in Fig.~\ref{CuK}d along the out-of-plane direction for fixed $H$ = 0.09 and $K$ = 0. For an unbiased comparison between the experimental and modeled dispersion [Figs.~\ref{CuK}e,f], we compute $\chi''(\textbf{q},\omega)$ of La$_{1.8}$Sr$_{0.2}$CuO$_4$ using the same parameters as in Ref.~\onlinecite{Hepting2022} for La$_{1.84}$Sr$_{0.16}$CuO$_4$, and only adjust the different hole-doping level $x$ = 0.2 in the calculation. Remarkably, Figs.~\ref{CuK}e,f reveal that both the in- and out-of-plane dispersion extracted from our experimental data essentially match with the computed dispersion, although the model appears to overestimate the plasmon energy. This small but systematic discrepancy will be discussed in more detail below. Nonetheless, the semiquantitative agreement between the experimental and the computed dispersion corroborates our identification of the plasmon mode in the Cu $K$-edge RIXS spectra, which in turn substantiates the emergence of plasmon quasiparticles in La$_{2-x}$Sr$_{x}$CuO$_4$ in the $T$-linear resistivity regime.

\subsection{Temperature dependence of the plasmon excitation}

\begin{figure}[tb]
\includegraphics[width=1.0\columnwidth]{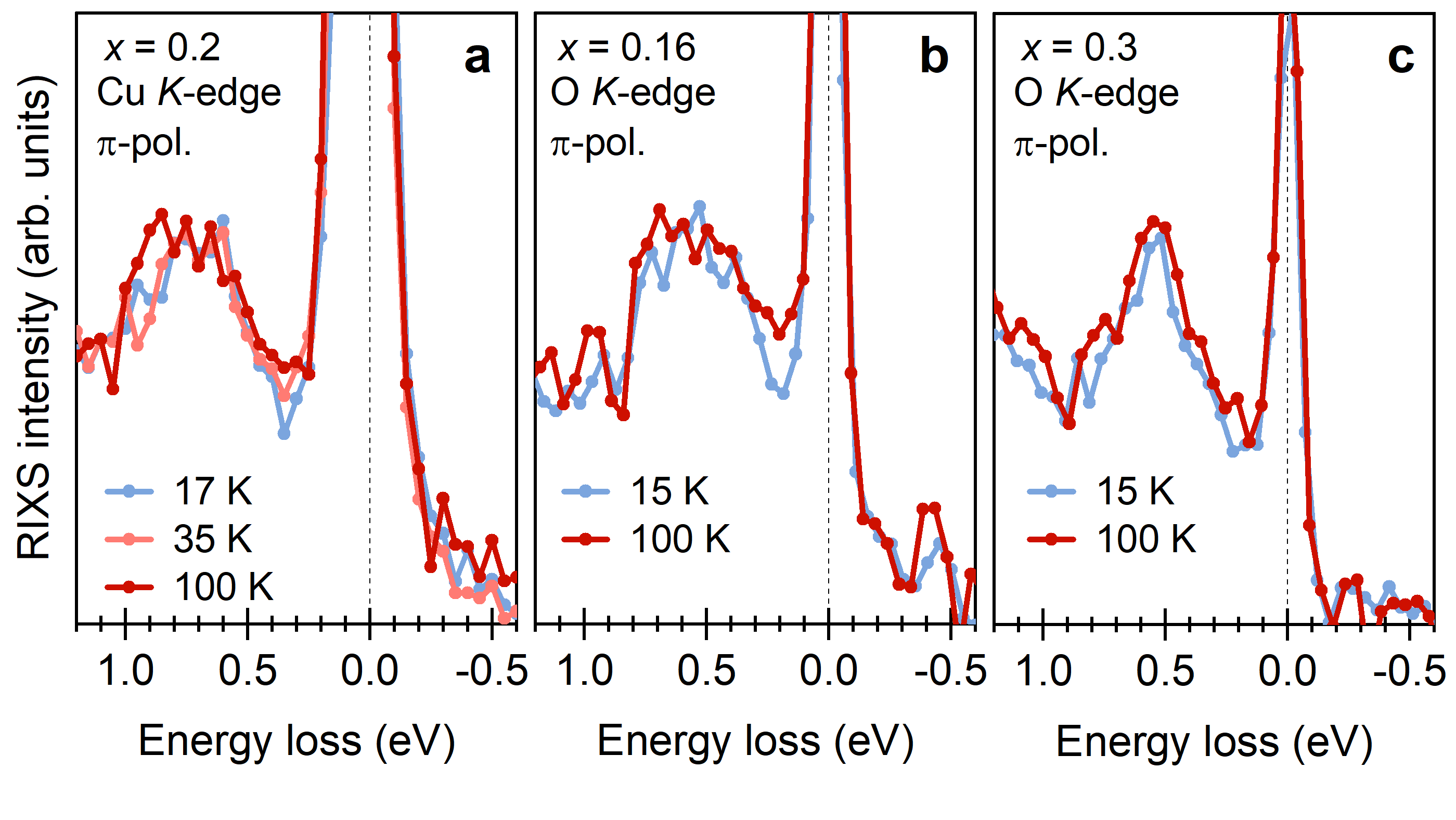}
\caption{Temperature dependence of the plasmon excitation. (a) Cu $K$-edge RIXS spectra of the La$_{1.8}$Sr$_{0.2}$CuO$_4$ crystal at $T$ = 17, 35, and 100 K, respectively. The momentum transfer is $H=K$ = 0.13 and $L^* = 6.5$. (b,c) O $K$-edge RIXS spectra of the (b) La$_{1.84}$Sr$_{0.16}$CuO$_4$ and (c) La$_{1.7}$Sr$_{0.3}$CuO$_4$ film at $T$ = 15 and 100 K, respectively. The momentum transfer is $H$ = 0.1, $K$ = 0, and $L^* = 0.5$. 
}
\label{LSCO_Tdep}
\end{figure}

In addition to the dependence of the plasmon excitation on the doping level $x$ at a fixed temperature [Fig.~\ref{LSCO_OK}], we investigate the temperature dependence of the plasmon for specific doping levels in the following. Figure~\ref{LSCO_Tdep}a displays the Cu $K$-edge RIXS spectra of the La$_{1.8}$Sr$_{0.2}$CuO$_4$ crystal at $T$ = 17, 35, and 100 K. Within the experimental error, the plasmon peak remains unchanged between 17 and 35 K, corresponding to measurements below and above the superconducting transition ($T_c\sim32$ K), respectively. Even an increase of the temperature to 100 K affects the plasmon excitation only marginally, that is, a subtle increase in the linewidth of the peak, likely due to thermal broadening of the collective excitation. To confirm the minor temperature dependence of the plasmon, we also perform O $K$-edge RIXS measurements on La$_{2-x}$Sr$_{x}$CuO$_4$ films with $x$ = 0.16 and 0.3 [Figs.~\ref{LSCO_Tdep}b,c], which exhibit a $T_c$ of 41 and 26 K, respectively. Similarly to Fig.~\ref{LSCO_Tdep}a, the temperature dependence of the plasmon peak in the latter two samples is small between 15 and 100 K and almost indiscernible from the  noise in the data, although a general trend of a linewidth broadening at higher temperatures appears evident.           

The observed absence of strong renormalization effects of the plasmon energy and linewidth across $T_c$ is consistent with detailed ellipsometry experiments of the optical plasmon branch in Bi-based cuprates in Ref.~\onlinecite{Levallois2016}. These experiments indicated that the temperature evolution of the optical plasmon branch with an energy of approximately 1 eV exhibits at most subtle anomalies at the superconducting transition. The detection of such subtle anomalies in the acoustic-like branches in Fig.~\ref{LSCO_Tdep} would require RIXS measurements with much higher statistics and denser temperature sampling intervals around $T_c$. 

Moreover, to detect significant renormalization effects of the acoustic-like plasmons across $T_c$, it may be necessary to conduct the RIXS measurements in immediate vicinity to the in-plane BZ center ($H=K \approx 0$). Specifically, at the in-plane BZ center the energy of acoustic-like plasmons becomes minimal, and can be comparable to the energy scale of the superconducting gap $2\Delta$, given that the interlayer-hopping $t_z$ of the material \cite{Hepting2022} is relatively small. Such proximity or a crossover of the two energy scales could affect the plasmon properties substantially. While  measurements in the present study were performed at momenta as small as $H$ = 0.04, the intense elastic line in the RIXS spectra progressively obscures the plasmon peak when approaching the in-plane BZ center. Thus, our present RIXS measurements do not provide sufficient information to ascertain or exclude a relationship between plasmons and superconductivity. Nevertheless, future RIXS investigations leveraging an improved energy resolution could yield new insights into this subject.

\subsection{Theoretical modeling of the doping dependence}

After our confirmation of the emergence of the plasmon excitation for $x$ = 0.2 by Cu $K$-edge RIXS [Fig.~\ref{CuK}] and our finding of a marginal temperature dependence [Fig.~\ref{LSCO_Tdep}], we return to the O $K$-edge RIXS data from Fig.~\ref{dopingtheo} for a detailed analysis of the doping-dependence in the framework of the $t$-$J$-$V$ model \cite{Greco2016,Greco2019,Greco2020,Nag2020,Hepting2022}. Specifically, Fig.~\ref{dopingtheo} compares the experimentally extracted plasmon energies (symbols) with the computed energies (dashed lines), which correspond to the maximum of the imaginary part of the charge susceptibility $\chi''(\textbf{q},\omega)$ calculated as a function of the doping $x$. The data are grouped according to the momenta $H$ = 0.04, 0.08, and 0.12, respectively, while $K = 0$ and $L^*=0.37$. Similarly to our calculation in Figs.~\ref{CuK}e,f, we employ the model parameters of La$_{1.84}$Sr$_{0.16}$CuO$_4$ derived in Ref.~\onlinecite{Hepting2022} (for details see Appendix~\ref{sec:theory}), and only adjust the doping level. For completeness, we also present the computed optical plasmon branch, and superimpose the optical plasmon energy of La$_{2-x}$Sr$_{x}$CuO$_4$ crystals (gray triangles) reported in optical reflectivity experiments \cite{Uchida1991}. Hence, these data points correspond to the plasmon energy at the three-dimensional BZ center ($H=K=L=0$), and were derived from the position of the peak in the loss function \cite{Uchida1991}. 

\begin{figure}[tb]
\includegraphics[width=1.0\columnwidth]{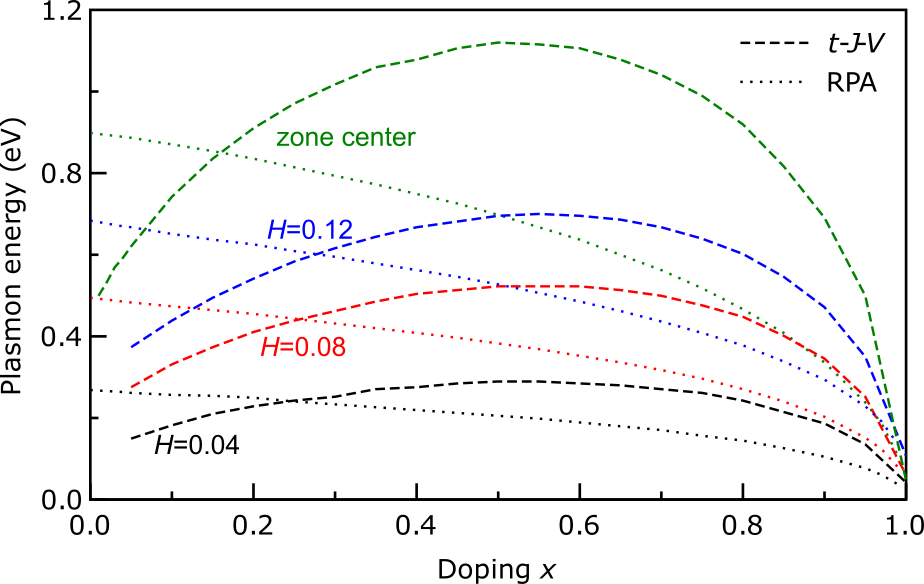}
\caption{Doping-dependence of the plasmon energy computed in the layered $t$-$J$-$V$ model (dashed lines) and the RPA (dotted lines). The parameters used in the RPA calculations are the same as those used for the $t$-$J$-$V$ model calculation, except for $V_c/t=14$, which was adjusted to match the experimental values of the optical plasma frequency ($\omega_{pl}^{opt}\sim 0.85$) at $x=0.16$ \cite{Suzuki1989,Uchida1991}. 
}
\label{tJVRPA}
\end{figure}

Notably, the computed plasmon energies in Fig.~\ref{dopingtheo} agree well with the experimentally observed energies for low dopings until approximately $x \approx 0.16$. This indicates that $t$-$J$-$V$ model calculations not only capture the behavior of plasmons in cuprates at a fixed doping level  \cite{Greco2016,Greco2019,Greco2020,Nag2020,Hepting2022}, but also describe their doping dependence without additional tuning of the parameters, at least for small $H$ and small dopings. The observed increase of the plasmon energy with increasing charge carrier concentration $x$ is intuitively expected for a system with some correspondence to a simple free electron model, where the plasmon energy scales with the charge carrier density. Yet, according to the basic plasmon theory \cite{mahan}, the plasmon energy should decrease with increasing hole doping, as it is proportional to $\sqrt{n/m^{*}}$, where $n$ is the electron density and $m^{*}$ the effective electron mass. Moreover, in RPA calculations (see Appendix~\ref{app:rpa} for details) the plasmon energy follows the band filling effect, \textit{i.e.}, the plasmon energy decreases with increasing hole-doping [Fig.~\ref{tJVRPA}]. Specifically, the plasmon energy computed in RPA is maximal for $x=0$, while it decreases with increasing $x$, and eventually vanishes for $x=1$, where the band becomes completely empty. Obviously, our observed increase of the plasmon energy in hole-doped La$_{2-x}$Sr$_{x}$CuO$_4$ until $x \approx 0.16$ is incompatible with the notion of the basic plasmon theory \cite{mahan} and the RPA results. A different theoretical calculation for La$_{2-x}$Sr$_{x}$CuO$_4$ \cite{Bauer2009}, which includes Hartree and exchange-correlation contributions, considered a mixed plasmon-phonon mode at very small in-plane momenta ($H \lesssim 0.005$). Whereas the plasmon-like branch of the mixed mode is essentially doping-independent between $x=0.156$ and $0.297$, the phonon-like branch increases in energy. However, the present RIXS data were acquired at much larger momenta which implies that such a coupling to phonons is not relevant for our plasmon dispersion in Fig.~\ref{dopingtheo}. In addition, Ref.~\onlinecite{Bauer2009} considered the case of an uncoupled plasmon, which shows an increase in energy for dopings between $x=0.156$ and $x=0.297$ for momenta $H \lesssim 0.005$, whereas it is almost doping-independent for slightly larger momenta. For a direct comparison to the present RIXS data, it will be interesting to carry out future calculations in analogy to Ref.~\onlinecite{Bauer2009}, but at larger momenta which coincide with the RIXS experiment and also for dopings below $x=0.156$.

In the present study, we employ the $t$-$J$-$V$ model to explore the entire doping range of La$_{2-x}$Sr$_{x}$CuO$_4$ at momenta that directly correspond to the RIXS experiment. The $t$-$J$-$V$ model takes into account strong correlations which are known to be responsible for the insulating state at half-filling ($x=0$) \cite{Lee2006}. Accordingly, Figs.~\ref{dopingtheo} and \ref{tJVRPA} illustrate that the plasmon energy computed in the $t$-$J$-$V$ model tends towards zero for $x=0$, which is consistent with the notion that insulators cannot exhibit plasmon excitations composed of conduction electrons (or holes). Note, however, that the plasmon energy at $x=0$ is expected to be non-zero also in the present calculations, because a finite exchange interaction $J$ induces a finite bandwidth even at $x=0$, if antiferromagnetic order is not considered \cite{Greco2016}. Nonetheless, the present theory is applicable at least down to $x=0.05$ [Figs.~\ref{dopingtheo} and \ref{tJVRPA}], which is the lowest hole-doping for which plasmons were detected in the experiment [Fig.~\ref{LSCO_OK}].     

Interestingly, above optimal doping and especially in the highly overdoped regime, the plasmon energies determined by RIXS tend to saturate around a doping that is lower than the prediction of the $t$-$J$-$V$ calculation, except for the momentum $H = 0.04$ [Fig.~\ref{dopingtheo}]. The optical plasmon energy determined from optical reflectivity experiments \cite{Uchida1991} even decreases between $x$ = 0.2 and 0.34, although we note that the decrease instead of a saturation might be due to possible sample quality issues for the highest doping in Ref.~\onlinecite{Uchida1991}. In general, the calculations suggest that the saturation trend is due to a broad peak structure of the plasmon energy as a function of doping [Fig.~\ref{tJVRPA}], which emerges naturally when considering that the plasmon energy tends towards zero for $x=0$ and 1. In particular, in the limit $x=1$, both the $t$-$J$-$V$ model calculations and the RPA results exhibit a strong decrease of the plasmon energy. This suggests that the $t$-$J$-$V$ model recovers the band filling effect when approaching the empty-band limit at $x=1$. 
However, Fig.~\ref{dopingtheo} indicates that the $t$-$J$-$V$ model calculation predicts a saturation of the plasmon energy at a doping level higher than that observed experimentally.

\section{Discussion and Conclusion}
\label{sec:discussion}

Our RIXS measurements at the O $K$-edge revealed that from very low dopings up to $x \approx 0.2$ the plasmon energy increases with increasing hole-doping [Fig.~\ref{dopingtheo}]. The increase in this doping regime is accurately captured by $t$-$J$-$V$ model calculations. The RPA calculations yield an opposite trend [Fig.~\ref{tJVRPA}], namely, a decrease of the plasmon energy with increasing hole-doping. This discrepancy points to the critical importance of an explicit inclusion of strong electron correlations into the modeling of plasmon excitations in cuprates, which is naturally implemented in the framework of the $t$-$J$-$V$ calculations. However, our study reveals that the $t$-$J$-$V$ model predicts a monotonic increase of the plasmon energy at least up to $x \approx 0.55$, whereas the experimental data indicates that the plasmon energy increase tends to saturate beyond $x \approx 0.2$ [Fig.~\ref{dopingtheo}]. 
In the following, we discuss three possible origins of this quantitative difference. 

First, we address potential experimental shortcomings and note that the doping dependence of the plasmon energy probed by RIXS is reminiscent of that of the optical plasmon energy observed by optics experiments \cite{Uchida1991,Kim2021}. This similarity rules out that the saturation is due to errors or inaccuracies of our RIXS measurements or data analysis. However, we cannot fully rule out that structural defects or chemical disorder of the dopant ions in our samples play a role in the saturation. In fact, the details of the distribution of the dopant ions \cite{Suyolcu2020,Suyolcu2022} as well as the emergence of oxygen vacancies \cite{Wang2022} and the effects of out-of-plane substitutional disorder
\cite{Narduzzo2008,Lee-Hone2018,Mahmood2019,Ozdemir2022} are currently under debate for cuprates. Along these lines, we note that the room-temperature resistivity of the $x$ = 0.4 film lies above that of the $x$ = 0.3 film [Fig.~\ref{transport}], whereas the resistivities decrease sequentially from $x$ = 0.05 to 0.3. This suggests that the electronic properties are affected by structural defects and/or disorder at least in our $x$ = 0.4 film. Yet, we stress that a Raman spectroscopy study reported that the formation of oxygen vacancies can be greatly alleviated in our ozone-assisted MBE grown La$_{2-x}$Sr$_{x}$CuO$_4$ films for dopings below $x$ = 0.35 \cite{Kim2017}. Moreover, the onset of the plasmon energy saturation is observed already between $x =0.16$ and 0.2, whereas the effects of disorder typically emerge only in the overdoped and highly overdoped regime \cite{Narduzzo2008,Lee-Hone2018,Mahmood2019,Ozdemir2022,Wang2022}. These considerations suggest that while disorder might have some effect on the propagation of the collective plasmon excitation for our overdoped samples, it cannot fully account for the observed plasmon energy saturation. 

Second, there might be an overestimation of correlations in the $t$-$J$-$V$ model, which pushes the crossover from a strongly correlated to a moderately or weakly correlated regime to higher dopings. In more detail, the broad maximum of the plasmon energy between $0.4 \lesssim x \lesssim 0.6$ in the $t$-$J$-$V$ calculation in Fig.~\ref{tJVRPA} can be regarded as a crossover between strong and weak correlation physics. At high dopings, electrons are more dilute and the local constraint in the $t$-$J$-$V$ model that prohibits double occupancy of electrons as well as the nearest-neighbor exchange interactions are less effective. This is the reason why the $t$-$J$-$V$ and RPA calculations show the same trend in the limit of $x = 1$ in Fig.~\ref{tJVRPA}. If the present $t$-$J$-$V$ calculations overestimate the correlation effect for high dopings in cuprates, a weaker correlation is expected to yield the broad peak at lower doping, which would be in better agreement with the experimental data in Fig.~\ref{dopingtheo}. Whether the experimental plasmon energy also exhibits a crossover behavior as predicted by the theory and hence starts to decrease for some doping level above $x = 0.4$ is an interesting topic for future experiments if La$_{2-x}$Sr$_x$CuO$_4$ films with such high dopings become available. 

Third, the activation of additional orbital degrees of freedom at large doping might play a significant role in the doping dependence. Such an activation is not captured by the one-band $t$-$J$-$V$ model, which is an effective model for the three-band Hubbard model, where the Cu $3d_{x^2-y^2}$ orbitals are hybridized with the 2$p_x$ and 2$p_y$ orbitals of the in-plane oxygen ions, forming the ZRS band \cite{Zhang1988}. Yet, there are theoretical indications that other planar \cite{Emery1987,Varma1997,Weber2008,Weber2012,Chen2013a,Ebrahimnejad2014,Ebrahimnejad2016,Adolphs2016,Kung2016} and non-planar orbitals \cite{Bianconi1988,Grilli1990,Feiner1992,Chen1992,Wang2010,Pavarini2001,Hozoi2011,Sakakibara2010,Sakakibara2012,Romberg1990,Jiang2020} should be taken into account in comprehensive models of cuprates. In the overdoped regime, specifically the Cu $3d_{3z^2-r^2}$, interstitial $4s$, and the $2p_{z}$ orbital of the apical oxygen were proposed to become relevant \cite{Feiner1992,Weber2010,Sakurai2011,Matt2018,Kramer2019}. For instance, a six-band Hubbard model, which includes the Cu $3d_{3z^2-r^2}$ and O $2p_{z}$ and $2p_{-z}$ orbitals, was employed to compute the doping dependence of the optical conductivity of La$_{2-x}$Sr$_x$CuO$_4$ \cite{Weber2010}. The obtained doping dependence is consistent with the experimentally observed continuous increase in optical spectral weight below 1.5 eV with increasing hole-doping \cite{Kim2021,Uchida1991}. Yet, the detailed dependence of the optical plasmon energy in the overdoped regime [Fig.~\ref{dopingtheo}] is difficult to discern in Ref.~\onlinecite{Uchida1991} and might be affected by disorder in the overdoped single-crystalline samples. On the other hand, a recent ellipsometry study on highly oxidized La$_{2-x}$Ca$_x$CuO$_4$ films \cite{Kim2021} carefully decomposed the optical spectral weight below 1.5 eV into a Drude peak and contributions from intra-ZRS transitions, revealing that the former component as well as the (bare) plasma frequency saturate above $x \approx 0.2$. By contrast, the second component associated with the intra-ZRS transitions and the Cu $3d_{3z^2-r^2}$ and O $2p_{z}$ orbitals increases continuously up to the highest measured doping. The sum of both components is again compatible with the spectral weight evolution calculated in Ref.~\onlinecite{Weber2010}. These distinct behaviors of the components in the optical spectra are strikingly reminiscent of our RIXS and XAS data. In particular, it seems likely that the peak in our RIXS spectra predominantly reflects the dynamics of the Drude charge carriers in the planar orbitals, whereas holes that are added to the system beyond $x \approx 0.2$ might be mostly associated with non-planar orbitals and do not contribute to the collective plasmon excitation probed by RIXS. On the other hand, the continuous increase of the XAS spectral weight of the hole-peak up to $x$ = 0.4 [Fig.~\ref{LSCO_OK}a-f] might reflect the total number of doped holes, both in the planar and non-planar orbitals. In this context, note that our XAS includes contributions from both planar and non-planar orbitals due to the employed measurement geometry (see Appendix~\ref{sec:experiment} for details). Hence, future XAS and RIXS experiments in a measurement geometry \cite{Chen1992} that is exclusively sensitive to charge carriers in the non-planar orbitals are highly desirable for confirming the above scenario and gaining further insights into the mobility of holes in the highly overdoped regime of cuprates.

Besides the observed saturation of the plasmon energy, our results provide fresh insights into the debate on whether plasmons can exist at all doping levels in the cuprate phase diagram, or if specific phases preclude their emergence. In particular, previous reflection EELS experiments on Bi$_{2.1}$Sr$_{1.9}$CaCu$_2$O$_{8+x}$ \cite{Husain2019,Mitrano2019} concluded that well-defined plasmon excitations are replaced by a featureless, temperature- and momentum-independent continuum in the strange metal regime, although the authors specified in a recent comment \cite{Husain2021} that they identified the spectral feature as a plasmon excitation for momenta $H \leq 0.12$. For larger momenta, it was suggested that the decay of the plasmon into a continuum is unusually strong \cite{Husain2019,Mitrano2019,Husain2021}, and that the putative quantum critical nature of the continuum can be captured by holographic theories \cite{Romero2019}. Yet, our combined O $K$- and Cu $K$-edge RIXS results on La$_{2-x}$Sr$_{x}$CuO$_4$ clearly indicate that dispersive plasmon excitations exist for dopings between $x$ = 0.05 and 0.4, including the strange metal regime at $x \sim 0.2$ for momenta at least up to $H$ = 0.15. Moreover, between $x$ = 0.05 and 0.2, the excitation energies are theoretically well described by the $t$-$J$-$V$ model without any fine-tuning of the parameters, which is a long-established model for cuprates. However, the linewidths of the plasmon excitations in Figs.~\ref{CuK}c,d are comparable to the respective plasmon energies for all measured momenta, indicating that the mode is appreciably damped. In general, several mechanisms can be responsible for a damping of plasmons in the long-wavelength limit around the BZ center, including scattering with phonons and Umklapp scattering \cite{Felde1989}, which can provide the momentum necessary to couple the plasmon to the electron-hole continuum. Furthermore, the damping around the BZ center depends on the detailed balance between the correlation strength and the size of the plasmon gap due to the interlayer hopping $t_z$ \cite{Zinni2023}. At large momenta, a pronounced plasmon decay is expected even in a conventional Lindhard model picture due to Landau damping when the plasmon traverses the energy scale of intra- and interband transitions \cite{Fink2021}. A definite distinction between the presence of these mechanisms and a quantum critical decay \cite{Romero2019} appears unfeasible based on the present RIXS data, whereas a comprehensive study of the plasmon linewidth evolution for both small and large momenta might give critical insights. In particular, while previous reflection EELS studies were focused on relatively large momenta \cite{Husain2019,Mitrano2019}, future Cu $K$-edge RIXS experiments are anticipated to provide detailed information about the entire BZ.

In a broader context, we point out that a cautious use of the terminology for the plasmon phenomenology in cuprates is important. For instance, Ref.~\onlinecite{Boyd2022} refers to the ensemble of branches between the optical plasmon and the lowest-energy (quasi-)acoustic branch in terms of a plasmon continuum. Such a continuum is not directly related to the decay of a nominally sharp plasmon excitation into a continuum due to various classical or quantum critical damping mechanisms \cite{Felde1989,Zinni2023,Fink2021,Romero2019}. In fact, the degree of the continuum-like character of the plasmon branch ensemble is determined by the total number $n$ of CuO$_2$ planes of a system, as the  number of (quasi-)acoustic branches is $n-1$ \cite{Bozovic1990}. Notably, for a given in-plane momentum, the signal probed by the reflection EELS technique is integrated over all (or at least a broad range of) out-of-plane momenta \cite{Boyd2022}. In contrast, the momentum resolution of RIXS is relatively high in all directions of reciprocal space \cite{Zhou2022}, which implies that especially for thin-film samples with a small number of CuO$_2$ planes the out-of-plane integration can be close to the limit of covering only a single branch. Whether this intrinsic difference between the two experimental techniques is responsible for the distinct spectral features observed below 1\,eV energy loss remains to be tested in future experiments. In particular, future Cu $K$-edge RIXS studies on Bi$_{2.1}$Sr$_{1.9}$CaCu$_2$O$_{8+x}$ and other hole-doped cuprates are highly desirable for establishing the universality of plasmon excitations in the strange metal regime of the cuprates and clarifying the nature of the plasmon decay.

\section*{Acknowledgments}
We thank A.V.~Boris, G.~Kim, and C.~Falter for fruitful discussions. A.G. acknowledges the Max-Planck-Institute for Solid State Research in Stuttgart for hospitality and financial support. H.Y. was supported by JSPS KAKENHI Grant No.~JP20H01856, Japan. Part of the research described in this paper was performed at the Canadian Light Source, a national research facility of the University of Saskatchewan, which is supported by the Canada Foundation for Innovation (CFI), the Natural Sciences and Engineering Research Council (NSERC), the National Research Council (NRC), the Canadian Institutes of Health Research (CIHR), the Government of Saskatchewan, and the University of Saskatchewan. This research used resources of the Advanced Photon Source, a U.S. Department of Energy (DOE) Office of Science user facility operated for the DOE Office of Science by Argonne National Laboratory under Contract No. DE-AC02-06CH11357.

\appendix

\section{Experimental details}
\label{sec:experiment}

A series of La$_{2-x}$Sr$_{x}$CuO$_4$ films with a thickness of 15 unit cells was grown by ozone-assisted molecular beam epitaxy (MBE) on (001) oriented LaSrAlO$_4$ (LSAO) substrates. The nominal Sr substitution levels were $x$ = 0.05, 0.1, 0.16, 0.2, 0.3, and 0.4. The hole-doping concentration in La$_{2-x}$Sr$_{x}$CuO$_4$ is directly proportional to the Sr-substitution level ($p = x$). The Van-der-Pauw method was used to determine the in-plane electrical resistivity. The $T_c$ values for $x$ = 0.1, 0.16, 0.2, 0.3, and 0.4 films are 35\,K, 41\,K, 32\,K, 26\,K, and below 2\,K, respectively. The $x$ = 0.05 film did not show an onset of a superconducting transition. In addition to the films, a single-crystal with a stoichiometry close to La$_{1.8}$Sr$_{0.2}$CuO$_4$ was grown by the optical floating zone technique. A standard Crystal Systems CSC FZ 1000 equipped with 300 W lamps was used and the crystal was grown at a rate of 1 mm/h in mixed Ar with 20\%  O$_2$ atmosphere at 3 bar. The refinement of single-crystal x-ray diffraction (XRD) data on two pieces broken off from the crystal indicated a Sr content of $x = 0.196 \pm 0.03$ and $0.17 \pm 0.03$, respectively. Energy dispersive x-ray spectroscopy (EDS) on the crystal surface yielded $x = 0.21$ when normalized to a nominally ideal Cu stoichiometry. The $T_c$ of the crystal was about 38\,K.

The doping dependence of the plasmon excitation in the La$_{2-x}$Sr$_{x}$CuO$_4$ films was measured using O $K$-edge RIXS at the REIXS beamline of the Canadian Light Source (CLS). The RIXS spectra were collected at 300 K using a Rowland circle spectrometer with a combined energy resolution $\Delta E \sim$ 190\,meV and linearly polarized photons ($\sigma$-polarization). The films were mounted such that the $c$- and the $a/b$-axes were lying in the scattering plane. The crystal structure of La$_{2-x}$Sr$_{x}$CuO$_4$ is orthorhombic, but as the difference between the $a$ and $b$ axes is small, and the films are twinned in the $ab$-plane, we do not distinguish between $a$ and $b$ in the following. The scattering angle $2\theta$ was kept fixed at $90^\circ$, while the angle $\theta$ between the sample surface and the incident x-rays was varied to probe different momentum transfers. The XAS data were taken with $\sigma$-polarized photons at an incident angle $\theta = 35^\circ$ in partial fluorescence yield using a silicon drift detector, collecting only the O $K_\alpha$ emission line.

We denote the momentum transfer by $(H,K,L)$ in reciprocal lattice units $(2\pi/a, 2\pi/b, 2\pi/c)$, where $a$, $b$, and $c$ are the lattice constants of La$_{2-x}$Sr$_{x}$CuO$_4$. Note that the crystallographic unit cell of La$_{2-x}$Sr$_{x}$CuO$_4$ contains two CuO$_2$ planes, whereas only the distance $d = c/2$ between adjacent CuO$_2$ planes is relevant for the periodicity of the plasmon dispersion \cite{Hepting2018}. Hence, the index $L^* = L/2$ is used in the following. Since the scattering angle $2\theta$ was fixed for the measurements at the REIXS beamline, each variation of the in-plane momentum transfer also leads to a (minor) change of the out-of-plane momentum. Specifically, a change of $H$ from 0.04 to 0.12 in the O $K$-edge RIXS measurements involves a concomitant change of $L^*$ from 0.393 to 0.346. However, since the plasmon dispersion as a function of $L^*$ is relatively shallow between $\sim$0.3 and 0.4 \cite{Greco2016}, we neglect the change of $L^*$ in the following and employ the averaged value of $L^* = 0.37$ for the analysis of the O $K$-edge RIXS data.  

Additional O $K$-edge RIXS measurements on selected La$_{2-x}$Sr$_{x}$CuO$_4$ films were conducted at the U41-PEAXIS beamline \cite{Lieutenant2016b,Schulz2020} of BESSY II at the Helmholtz-Zentrum Berlin (HZB). The spectra were collected at various temperatures between 15 and 100 K, with a combined energy resolution $\Delta E \sim$ 90\,meV. Typically, $\sigma$-polarized photons are used for soft x-ray measurements of plasmon excitations in cuprates \cite{Hepting2018,Nag2020,Lin2019,Hepting2022}, as the charge signal is enhanced over the magnetic signal. However, since only $\pi$-polarization was available for our RIXS experiment at U41-PEAXIS, the temperature-series was taken at momenta that correspond to incident angles $\theta$ close to normal incidence. Accordingly, the projection of the polarization vector of the electric field of the $\pi$-polarized photons onto the CuO$_2$ planes was maximized, enhancing the charge signal as much as possible. The scattering angle $2\theta$ at U41-PEAXIS can be varied across a broad range. For our measurement, $2\theta$ was approximately $137^\circ$. The corresponding in-plane momentum transfer was $H$ = 0.1 and the out-of-plane momentum transfer $L^* = 0.5$.  

The Cu $K$-edge RIXS measurements on the La$_{1.8}$Sr$_{0.2}$CuO$_4$ single-crystal were carried out at the MERIX spectrometer at sector 27 at the Advanced Photon Source (APS). RIXS spectra were collected between 17 and 100\,K. The incident photons were monochromatized by a four-bounce monochromator with asymmetrically cut Si (4, 0, 0) crystals. A spherical diced Ge (3, 3, 7) analyzer was used and the flight path between the analyzer and sample/detector was approximately 1 meter. The combined energy resolution was $\Delta E \sim$ 80\,meV. The incident angle $\theta$ and the scattering angle $2\theta$ could be varied independently from each other. To minimize the strong elastic scattering signal in the RIXS spectra, $\pi$-polarized photons and a scattering angle $2\theta$ close to $90^\circ$ was chosen. Specifically, for all measured $H$, $K$, and $L$ values of momentum-dependence of the plasmon excitation, the angle $2\theta$ remained close to $90^\circ$. The scattering geometry of the RIXS experiment was such that the $c$-axis and the diagonal between the $a$- and $b$-axis of La$_{1.8}$Sr$_{0.2}$CuO$_4$ were lying in the scattering plane. The XAS data were collected in the fluorescence yield mode with $\pi$- and $\sigma$-polarized photons, respectively, at a shallow incident angle $\theta$ (close to grazing incidence). Hence, the polarization vector of the electric field of the incoming photons lies within the $a/b$ plane of the crystallographic unit cell for $\sigma$-linearly polarized photons, whereas it is mostly parallel to the $c$-axis for $\pi$-polarized photons, but with a small projection on the $a/b$ plane due to a small $\theta$.

\section{RIXS raw data and fits}
\label{app:fit}

\begin{figure}[tb]
\includegraphics[width=.75\columnwidth]{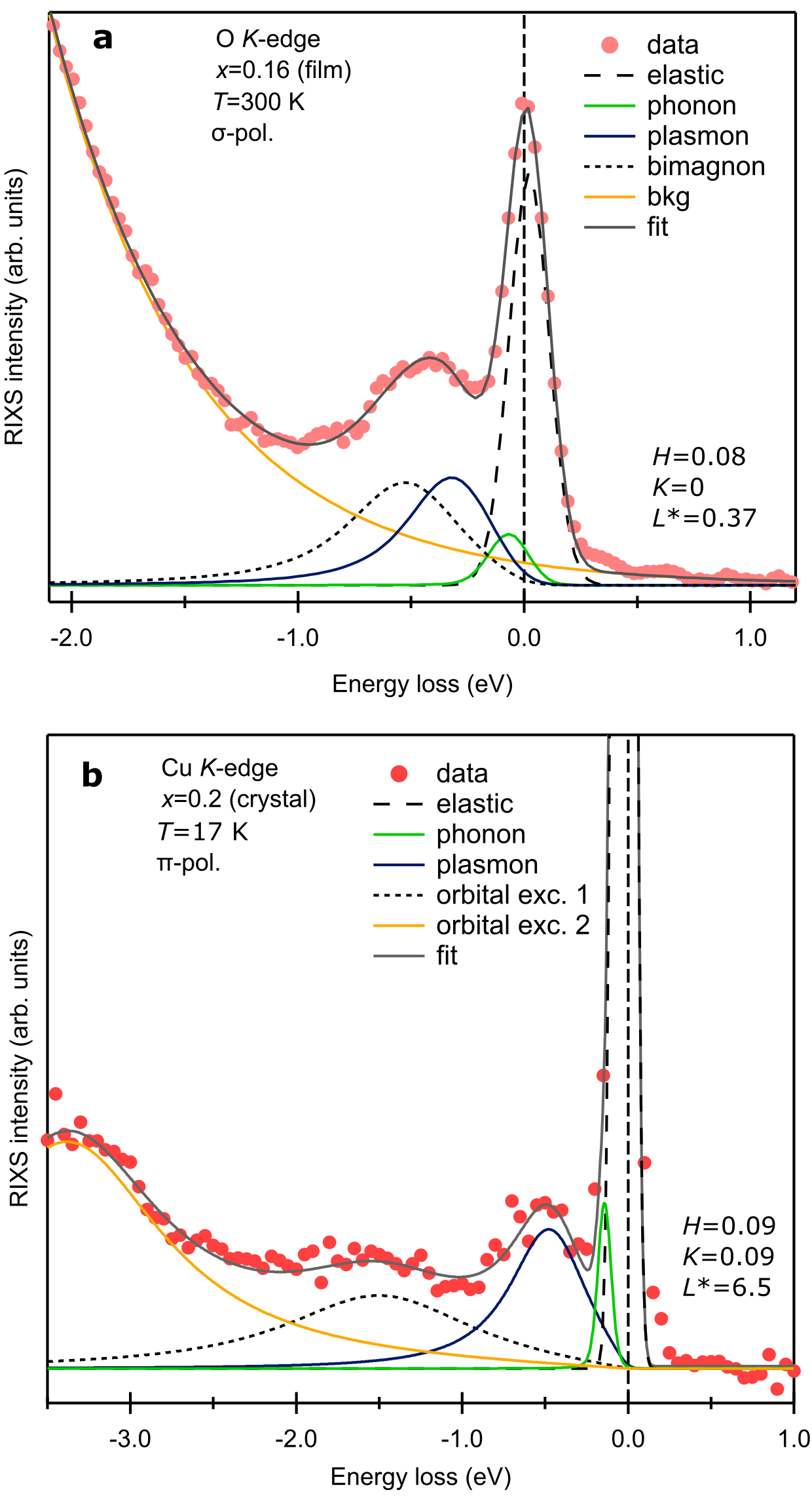}
\caption{Representative fits of the RIXS spectra. (a) Fit of the O $K$-edge RIXS spectrum of the $x$ = 0.16 film taken at momentum $H$ = 0.08, $K$ = 0, and $L^*$ = 0.37. The individual contributions of the fit are described in the text. (b) Fit of the Cu $K$-edge RIXS spectrum of the $x$ = 0.2 single-crystal taken at momentum momentum $H$ = 0.09, $K$ = 0.09, and $L^*$ = 6.5. 
}
\label{fits}
\end{figure}

\begin{figure*}[tb]
\includegraphics[width=2.0\columnwidth]{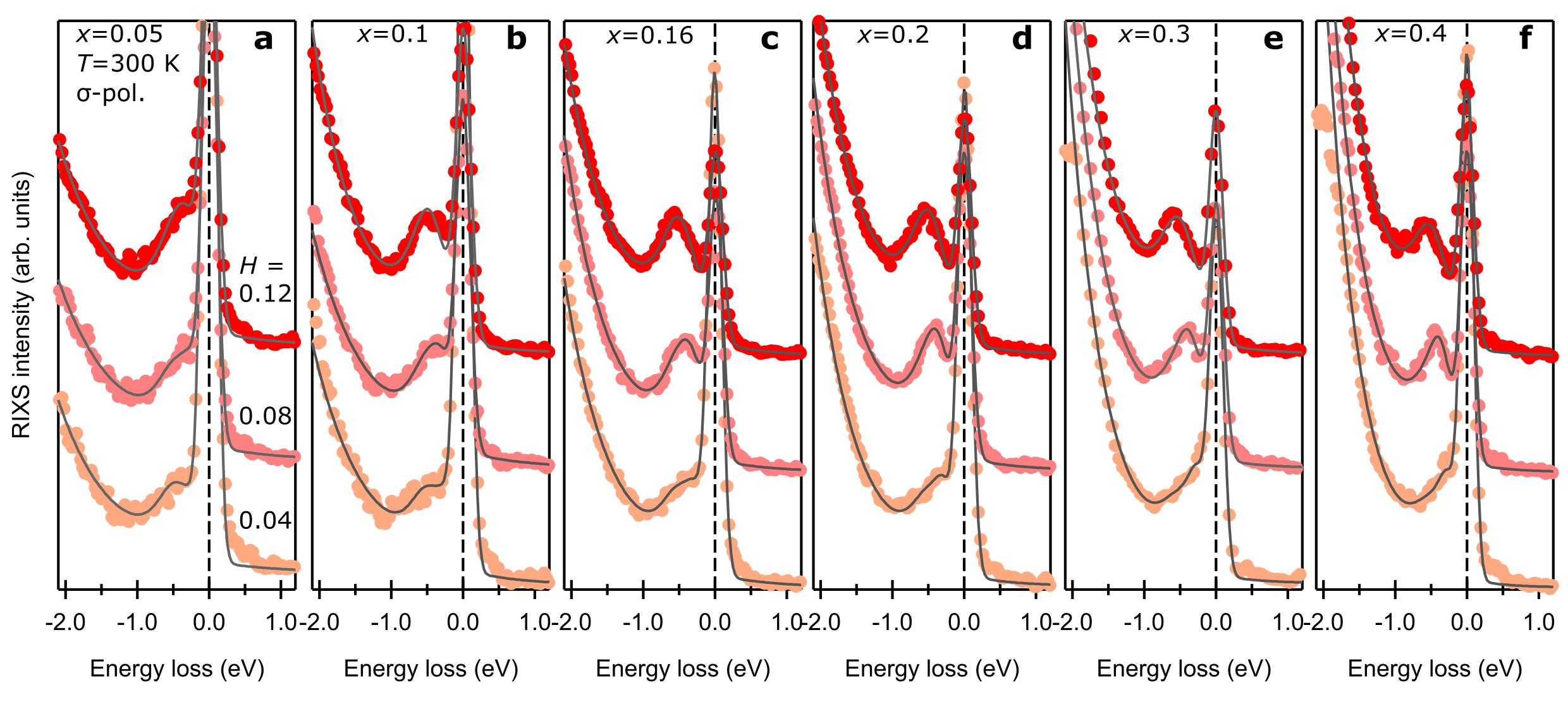}
\caption{(a)-(f) Raw RIXS spectra (filled symbols) of the La$_{2-x}$Sr$_{x}$CuO$_4$ films with various hole-doping levels $x$ taken at the O $K$-edge, together with fits (gray lines). Curves for different momenta are offset in the vertical direction for clarity. 
}
\label{LSCO_OK_RAW}
\end{figure*}

Figure~\ref{fits}a shows all components of a representative fit of the O $K$-edge RIXS spectra of the La$_{2-x}$Sr$_x$CuO$_4$ films of Fig.~2 of the main text. The displayed RIXS spectrum was taken on the $x$ = 0.16 film at momentum $H$ = 0.08, $K$ = 0, and $L^*$ = 0.37. The elastic peak was modeled by a Gaussian and the other contributions in the spectra by anti-symmetrized Lorentzians \cite{Hepting2018,Nag2020}, convoluted with the energy resolution of $\Delta E$ = 190\,meV via Gaussian convolution. The anti-symmetrized Lorentzian profiles ensure zero intensity at zero energy loss (prior to convolution) for the inelastic features. The inelastic features are assigned to a phonon, a plasmon, and a non-dispersive bimagnon. In the fits, the energy of the bimagnon peaks was kept fixed, while all other fit parameters were varied. Furthermore, the spectra contain a broad and intense background (bkg) feature, which peaks beyond 2 eV energy loss. This feature consists of fluorescence and $dd$-excitations \cite{Bisogni2012a}. In the fits, we included the tail of the feature below 2.1 eV energy loss. For clarity, the signal originating from this feature was subtracted in the RIXS spectra shown in Fig.~2 of the main text. A collection of the raw data of the O $K$-edge RIXS spectra of the La$_{2-x}$Sr$_x$CuO$_4$ films is shown in Fig.~\ref{LSCO_OK_RAW}.   

\begin{figure}[tb]
\includegraphics[width=1.\columnwidth]{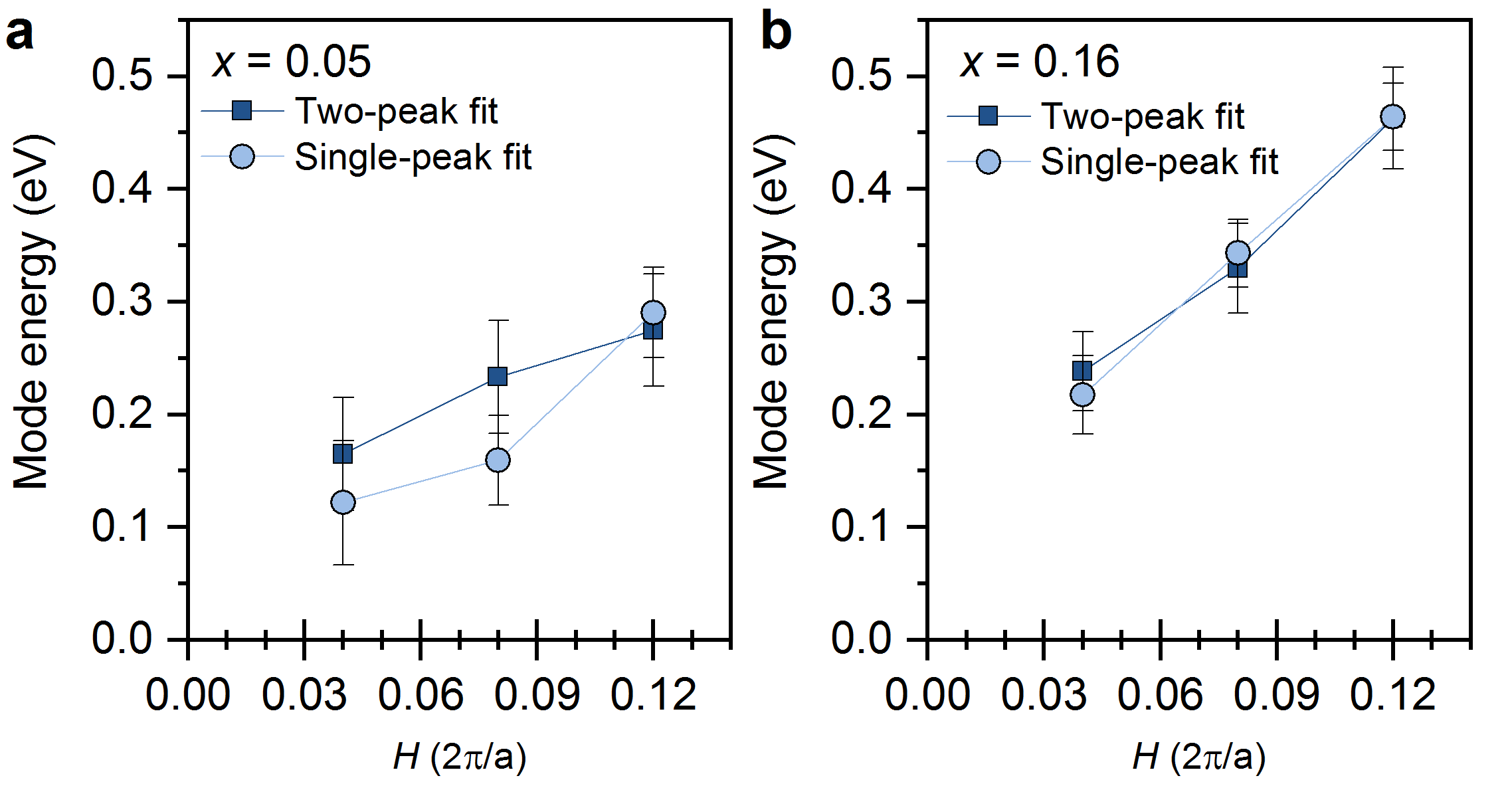}
\caption{Results for alternative fits of the RIXS spectra of the La$_{2-x}$Sr$_{x}$CuO$_4$ films. (a) Fit results for the $x = 0.05$ film. The blue squares (two-peak fit) indicate the plasmon energy extracted from a fit that includes separate plasmon and bimagnon peaks, in analogy to Fig.~\ref{fits}a. The light blue circles (single-peak fit) indicate the energy of a single peak that comprises both the plasmon and the bimagnon feature (fit not shown here). (b) Fit results for the $x = 0.16$ film.   
}
\label{LSCO_one_peak}
\end{figure}

The observed trend of an increase of the peak energy with increasing doping concentration up to $x = 0.16$ does not depend on the details of our fitting procedure, such as the exact treatment of the bimagnon peak. For instance, a fit that captures both the plasmon and the bimagnon feature by a single broad peak function yields similar results to Fig.~3 of the main text. This is illustrated in Fig.~\ref{LSCO_one_peak}, which shows the energy of the plasmon peak (blue squares, two-peak fit) extracted from a fit that includes separate plasmon and bimagnon peaks (see also Fig.~3 of the main text), together with the energy of a single peak (light blue circles, single-peak fit) that comprises both the plasmon and the bimagnon feature.

\begin{figure}[tb]
\includegraphics[width=.95\columnwidth]{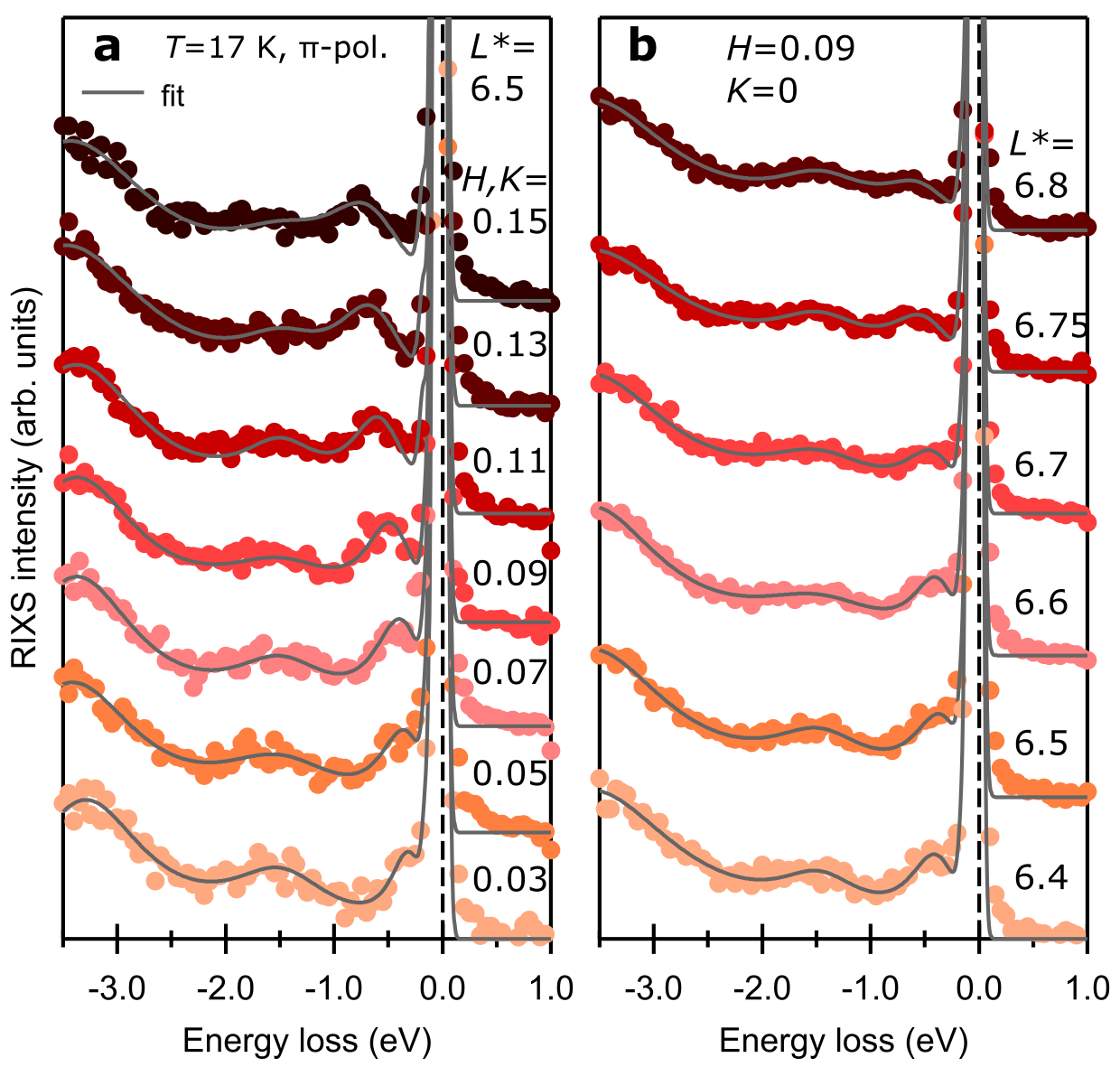}
\caption{Raw RIXS spectra (filled symbols) of the La$_{1.8}$Sr$_{0.2}$CuO$_4$ single-crystal taken at the Cu $K$-edge, together with fits (gray lines). Curves for different momenta are offset in the vertical direction for clarity. (a) RIXS spectra for various in-plane momenta along the ($H,H$) direction, while the out-of-plane momentum is fixed to $L^* = 6.5$. (b) RIXS spectra for momenta along the $L^*$ direction, while $H=0.09$ and $K=0$ are fixed.  
}
\label{CuK_RAW}
\end{figure}

The components of a representative fit of the Cu $K$-edge RIXS spectra of the La$_{1.8}$Sr$_{0.2}$CuO$_4$ single-crystal of Fig.~4 of the main text is given in Fig.~\ref{fits}b. The displayed RIXS spectrum was taken at momentum $H$ = $K$ = 0.09, and $L^*$ = 6.5. The elastic peak was modeled by a Gaussian and the other contributions in the spectra by anti-symmetrized Lorentzians, convoluted with the energy resolution of $\Delta E$ = 80\,meV via Gaussian convolution. The inelastic features are assigned to a phonon, a plasmon, a non-dispersive orbital excitation around 1.5~eV energy loss (orbital exc. 1), and an additional orbital excitation at higher energy losses (orbital exc. 2)  \cite{Ellis2011}. A collection of the raw data of the Cu $K$-edge RIXS spectra is shown in Fig.~\ref{CuK_RAW}.

\section{Comparison to literature}
\label{app:literature}

\begin{figure}[tb]
\includegraphics[width=1.0\columnwidth]{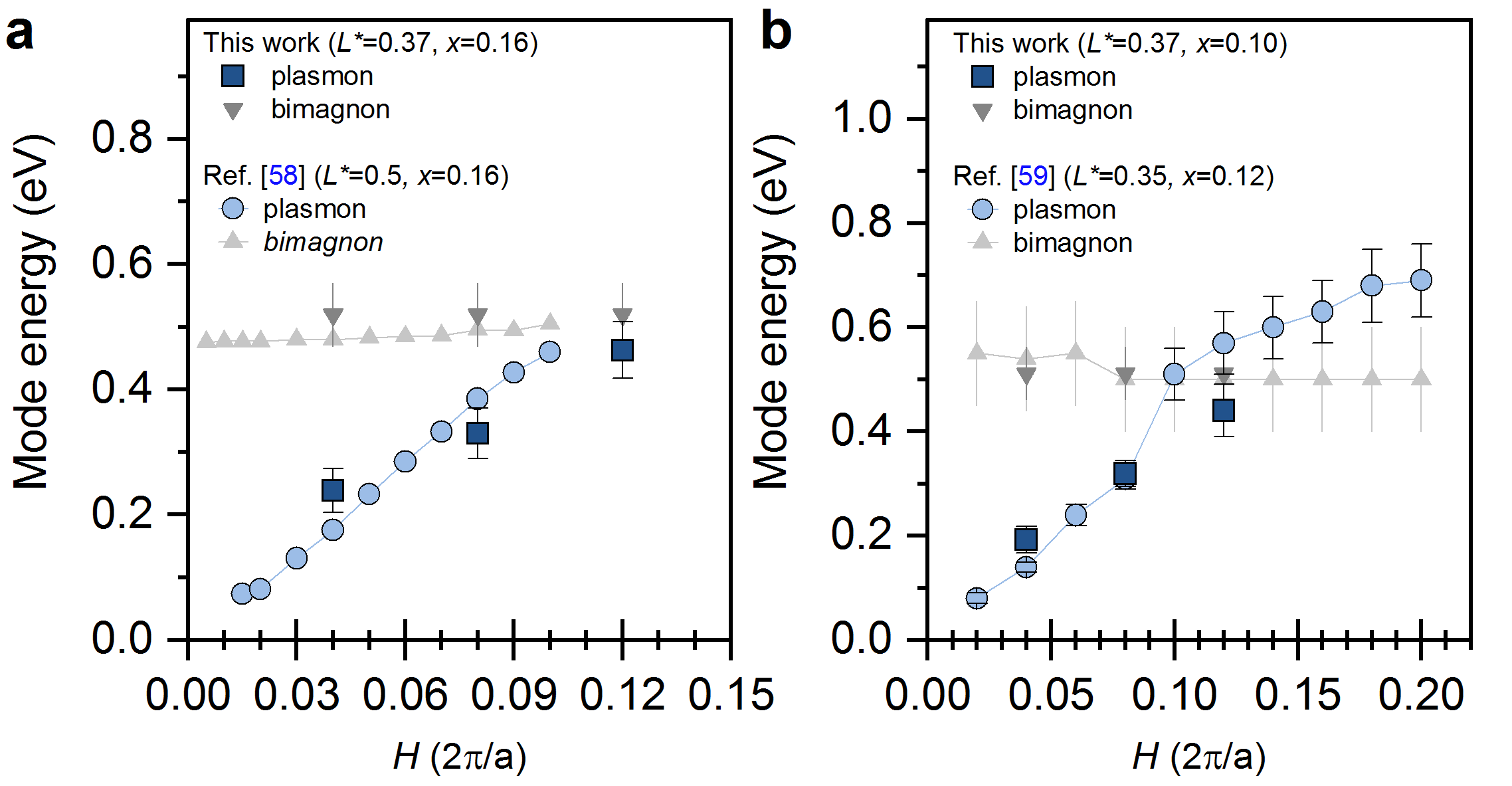}
\caption{Comparison of the fit results of the RIXS spectra with literature. (a) Blue squares (dark gray triangles) are the plasmon (bimagnon) energies extracted from the O $K$-edge RIXS spectra of the La$_{2-x}$Sr$_x$CuO$_4$ film with $x = 0.16$. Blue circles (light gray triangles) are the plasmon (bimagnon) energies of a La$_{2-x}$Sr$_x$CuO$_4$ single-crystal with $x = 0.16$ from Ref.~\onlinecite{Nag2020}. (b) Fit results of the La$_{2-x}$Sr$_x$CuO$_4$ film with $x = 0.1$ and the results of Ref.~\onlinecite{Singh2022} on a La$_{2-x}$Sr$_x$CuO$_4$ single-crystal with $x = 0.12$.
}
\label{LSCO_Hdep}
\end{figure}

Figure~\ref{LSCO_Hdep} shows the plasmon (bimagnon) energies of two representative La$_{2-x}$Sr$_x$CuO$_4$ film ($x = 0.1$ and 0.16) as a function of the momentum transfer $H$. In addition, the plasmon (bimagnon) energies taken at similar momenta and dopings in Refs.~\cite{Nag2020,Singh2022} are superimposed. The plasmon (bimagnon) energies determined in the present work are similar to those from Refs.~\cite{Nag2020,Singh2022}.

\section{Theoretical details}
\label{sec:theory}

For the theoretical description of plasmon excitations in cuprates, we employ the layered $t$-$J$-$V$ model \cite{Greco2016,Greco2019}, which incorporates the three-dimensional character of the plasmons \cite{Hepting2018} via the long-range (interlayer) Coulomb interaction $V$:

\begin{multline}
H = -\sum_{i, j,\sigma} t_{i j}\tilde{c}^\dag_{i\sigma}\tilde{c}_{j\sigma} + 
\sum_{\langle i,j \rangle} J_{ij} \left( \vec{S}_i \cdot \vec{S}_j - \frac{1}{4} n_i n_j \right) \\
+ \frac{1}{2}\sum_{i,j} V_{ij} n_i n_j \, 
\label{tJV}
\end{multline}
\noindent where the sites $i$ and $j$ run over a three-dimensional lattice. The hopping $t_{i j}$ takes a value $t$ $(t')$ between the first (second) nearest-neighbors on the square lattice. The hopping integral between layers is scaled by $t_z$. $\langle i,j \rangle$ denotes a nearest-neighbor pair of sites. The exchange interaction $J_{i j}=J$ is considered only inside the plane; the exchange term between the planes ($J_\perp$) is much smaller than $J$ \cite{Thio1988}. $\tilde{c}^\dag_{i\sigma}$ ($\tilde{c}_{i\sigma}$) is the creation (annihilation) operators of electrons with spin $\sigma(=\uparrow, \downarrow)$ in the Fock space without any double occupancy. $n_i=\sum_{\sigma} \tilde{c}^\dag_{i\sigma}\tilde{c}_{i\sigma}$ is the electron density operator and $\vec{S}_i$ is the spin operator. $V_{ij}$ is the long-range Coulomb interaction on the lattice and is given in momentum space by \cite{Becca1996}:

\begin{equation}
V({\bf q})=\frac{V_c}{A(q_x,q_y) - \cos q_z} \,,
\label{LRC}
\end{equation}

\noindent where $V_c= e^2 d(2 \epsilon_{\perp} a^2)^{-1}$ and 
\begin{equation}
A(q_x,q_y)=\alpha (2 - \cos q_x - \cos q_y)+1 \,.
\end{equation}

\noindent Here $\alpha=\frac{\tilde{\epsilon}}{(a/d)^2}$, where $\tilde{\epsilon}=\epsilon_\parallel/\epsilon_\perp$,  
and $\epsilon_\parallel$ and $\epsilon_\perp$ are the dielectric constants parallel and perpendicular to the planes, respectively; $e$ is the electric charge of electrons;  
$a$ is the lattice spacing in the planes and the in-plane momentum $q_\parallel=(q_x,q_y)$ is given in units of $1/a$; similarly $d$ is the distance between the planes and the out-of-plane momentum $q_z$ is given in units of $1/d$.

To compute the plasmon excitations we use the large-$N$ formalism given in Ref.~\onlinecite{Greco2016}. The charge-charge correlation function can be calculated as

\begin{equation}\label{CH}
\chi({\bf q},\mathrm{i}\omega_n)= N \left ( \frac{x}{2} \right )^{2} D_{11}({\bf q},\mathrm{i}\omega_n)\,.
\end{equation}

\noindent Here $D_{11}$ is the element $(1,1)$ of the $6 \times 6$ bosonic propagator 

\begin{eqnarray}
[D_{ab}({\bf q},\mathrm{i}\omega_n)]^{-1}
= [D^{(0)}_{ab}({\bf q},\mathrm{i}\omega_n)]^{-1} - \Pi_{ab}({\bf q},\mathrm{i}\omega_n)\,,
\label{dyson}
\end{eqnarray}

\noindent  where $\omega_n$ are the Matsubara frequencies and $a$,$b$ run from $1$ to $6$. The factor $N$ in Eq.~(\ref{CH}) comes from the sum over the $N$ fermionic channels after the extension of the spin index from $2$ to $N$. The corresponding expressions for the bare bosonic propagator   $D^{(0)}_{ab}({\bf q},\mathrm{i}\omega_n)$ 
and the bosonic self-energy $\Pi_{ab}({\bf q},\mathrm{i}\omega_n)$ are given in Ref.~\onlinecite{Greco2016}.

The large-$N$ formalism successfully captured the experimentally observed low-energy plasmon dispersions in various cuprates for specific doping levels in prior works \cite{Greco2019,Greco2020,Nag2020,Hepting2022}. As also used in Ref.~\onlinecite{Hepting2022} to fit and reproduce the experimental plasmon dispersion in La$_{1.84}$Sr$_{0.16}$CuO$_4$, we use the parameters $t/2=0.35$ eV, $t'/t=-0.2$, $t_z/t=0.01$, $J/t=0.3$, $V_c/t=31$, and $\alpha=3.5$ for La$_{2-x}$Sr$_{x}$CuO$_4$ in the present work. The number of planes is $N_z=30$ and temperature $T=0$.

\section{Summary of the RPA formalism}
\label{app:rpa}

In the RPA the charge correlation function is given by \cite{mahan}

\begin{eqnarray}
\chi_{\rm RPA}({\bf q},\mathrm{i}\omega_{n})=\frac{\chi^{(0)}({\bf q},\mathrm{i}\omega_{n})}
{1-V({\bf q})\chi^{(0)}({\bf q},\mathrm{i}\omega_{n})}, 
\end{eqnarray}
\noindent  where $\chi^{(0)}({\bf q},\mathrm{i}\omega_{n})$ is the 
Lindhard function which accounts for the particle-hole continuum.
The electron dispersion $E_{\bf {k}}$ is
\begin{eqnarray}
E_{\bf{k}} = E_{\bf{k}}^{\parallel}  + E_{\bf{k}}^{\perp},
\label{Ek}
\end{eqnarray}

\noindent where 
\begin{eqnarray}
E_{\bf{k}}^{\parallel} =-2 t(\cos k_{x}+\cos k_{y})
-4t'\cos k_{x} \cos k_{y} - \mu  \,, \label{EparaRPA} \\
E_{\bf{k}}^{\perp} = -2 t_{z} (\cos k_x-\cos k_y)^2 \cos k_{z}  \,. \label{EperpRPA}
\end{eqnarray}

\bibliography{cuprates}
\end{document}